\documentclass[aps,superscriptaddress,showpacs,nofootinbib,eqsecnum]{revtex4}

\usepackage{epsfig}

\input{epsf}
\begin{document}

\title{Asymptotically Improved Convergence of  
Optimized Perturbation Theory in the Bose-Einstein Condensation Problem}

\author{Jean-Lo\"{\i}c Kneur}
\email{kneur@lpm.univ-montp2.fr}
\affiliation{Laboratoire de Physique Math\'{e}matique et Th\'{e}orique - CNRS - UMR
5825 Universit\'{e} Montpellier II, France}

\author{Marcus B. Pinto}
\email{marcus@lpm.univ-montp2.fr}\thanks{$^1$ Permanent address}
\affiliation{Laboratoire de Physique Math\'{e}matique et Th\'{e}orique - CNRS - UMR
5825 Universit\'{e} Montpellier II, France}
\affiliation{Departamento de F\'{\i}sica,
Universidade Federal de Santa Catarina,
88040-900 Florian\'{o}polis, SC, Brazil}

\author{Rudnei O. Ramos}
\email{rudnei@dft.if.uerj.br}
\affiliation{Departamento de F\'{\i}sica Te\'orica,
Universidade do Estado do Rio de Janeiro,
20550-013 Rio de Janeiro, RJ, Brazil}

\begin{abstract}

We investigate the convergence properties of optimized perturbation
theory, or linear $\delta$ expansion (LDE), within the context of finite
temperature phase transitions. Our results prove the reliability of
these methods, recently employed in the determination of the critical
temperature $T_c$ for a system of weakly interacting homogeneous dilute
Bose gas. We carry out the explicit LDE optimized calculations and also
the infrared analysis of the relevant quantities involved in the
determination of $T_c$ in the large-$N$ limit, when the relevant
effective static action describing the system is extended to $O(N)$
symmetry. Then, using an efficient resummation method, we show how the
LDE can exactly reproduce the known large-$N$ result for $T_c$ already at
the first non-trivial order. Next, we consider the finite $N=2$ case
where, using similar resummation techniques, we improve the analytical
results for the nonperturbative terms involved in the expression for the
critical temperature allowing comparison with recent Monte Carlo
estimates of them. To illustrate the method we have considered a simple
geometric series showing how the procedure as a whole works consistently
in a general case.

\end{abstract}


\centerline{\large \it In Press Physical Review A (2003) }

\maketitle

\section{INTRODUCTION}

Scalar field theories are extremely important in the study of symmetry
breaking and restoration in different branches of physics such as
cosmology, particle physics and condensed matter physics, where they may
represent inflatons, Higgs particles, quark--anti-quark bound states,
Cooper pairs, bosonic atoms and molecules, respectively. In most cases
the vacuum expectation value of those scalar fields represents an order
parameter that signals phase transitions associated to
symmetry breaking/restoration \cite{goldenfeld}. 

In general, one important problem we have to deal with when studying
phase transitions in field theory regards the reliability of
perturbation theory and its eventual breakdown. In this case, a
nontrivial problem arises since nonperturbative methods must be used.
This is the case in those physical situations involving a second order,
or weakly first order phase transition, where we have to consider the
problem of infrared (IR) divergences that become progressively more
important as one approaches the critical temperature, from above or
below, and that will unavoidably spoil any perturbative attempt to
compute relevant quantities there. In those situations we must find
appropriate methods to take into account the large IR corrections,
present in the form of large field fluctuations. There is a variety of
nonperturbative methods that can be used in order to account for these
corrections, including the recent dynamical Boltzmann-like approach that
deals directly with the large field fluctuations \cite{subcrit}. At the
same time, the most common methods, at equilibrium, try to resum the
leading IR corrections. This happens for example, through the use of
$\varepsilon$-expansion techniques in order to compute corrections to
the critical exponents that control the singular behavior of physical
quantities near the critical point (as it is familiar from the theory of
critical phenomena \cite{critexp}), the large-$N$ method and other
approaches (for a review, see Ref. \cite{justin}). 

An issue that has attracted considerable attention recently and
that is associated to the perturbation theory breakdown problem is the
study of how interactions alter the critical temperature ($T_c$) of
Bose-Einstein condensation (BEC). Due to its nonperturbative nature,
this is clearly a non-trivial problem. On the other hand, studies
related to the BEC $T_c$ problem are particularly important nowadays due
the recent experimental realization of BEC on dilute atomic gases (for
reviews, see for instance, Ref. \cite{reviews}). The experimental
achievement of BEC has induced many theoretical investigations which
make use of methods developed to treat finite temperature quantum field
theories. At the same time, due to the high experimental precision with
which the parameters may be tuned, BEC experiments provide an important
laboratory to test many methods as well as models developed to treat
those theories (see, {\it e.g.}, Ref. \cite{prl}).

The studies concerning the equilibrium properties of BEC can be
addressed by means of a non-relativistic effective theory described by a
complex scalar field. In the dilute limit, which is the regime involved
in those experiments, only two body interactions are important
\cite{reviews} and one may then consider the following $U(1)$ invariant
finite temperature Euclidean action

\begin{equation} S_E = \int_0^\beta d\tau \int d^3 x \left\{
\psi^*({\bf x},\tau)\left( \frac{d}{d\tau}-\frac{1}{2m}\nabla^2\right)
\psi ({\bf x},\tau) -\mu \psi^* ({\bf x},\tau) \psi ({\bf x},\tau) +
\frac{2 \pi a}{m} \left[\psi^*({\bf x},\tau)\psi({\bf x},\tau) \right]^2
\right\}\;, 
\label{SE} 
\end{equation} 
where, in natural units, $\beta$ is the inverse of
the temperature, $\mu$ is the chemical potential
and $m$ represents the mass of the atoms. At the relevant low
temperatures involved in BEC the internal degrees of freedom are
unimportant and this can be taken as an effective model of hard core
spheres with local interactions for which $a$ represents the $s$-wave
scattering length.

The field $\psi$ can be decomposed into imaginary-time frequency modes
$\psi_j ({\bf x},\omega_j)$, with discrete bosonic Matsubara frequencies
$\omega_j = 2 \pi j/ \beta$, where $j$ is an integer. Near the
transition the chemical potential becomes very small as compared to the
temperature ($|\mu|<<T$) and, since the correlation length tends to
infinity, the distances are large compared to the thermal wavelength
$\lambda= \sqrt{2\pi\beta/m}$. Therefore, the non-zero Matsubara modes
decouple and one is left with an effective action for the field zero
modes ($j=0$) given by \cite {baymprl}

\begin{equation}
S_{3d} = \beta \int d^3 x \left\{
\psi_0^*\left(-\frac{1}{2m}\nabla^2- \mu \right )
\psi_0 + \frac{2 \pi a}{m}  \left[\psi_0^* \psi_0
\right]^2 \right\}\;,
\label{3d}
\end{equation}
where $\psi_0$ stands for the field's zero mode. Recently, Arnold, Moore
and Tom\'{a}sik \cite {second} have argued that when naively going from
the original action ($S_{E}$) to the reduced action ($S_{3d}$) by
ignoring the effects of non-zero frequency modes one misses the effects
that short-distances and/or high-frequency modes have on long-distance
physics. {}For the critical temperature of condensation as a function of
the density ($T_c(n)$), at second order, these effects can be absorbed
into a modification of the strengths of the relevant interactions which
means that one should consider the more general form for the reduced
effective action Eq. (\ref{3d})

\begin{equation}
S_{\rm eff} [\psi_0,\psi^*_0] = \beta \int d^3 x \left\{\psi^*_0 \left(
-{\cal Z}_{\psi} \frac{1}{2m}\nabla^2 -\mu_{3} \right)
\psi_0
+{\cal Z}_a \frac{2 \pi a}{m}  \left[\psi^*_0\psi_0
\right]^2 + {\cal O}\left[ \psi^*_0 \psi_0 |\nabla\psi|^2, (\psi^*\psi)^3\right]
\right\} + \beta F_{\rm vacuum}\;,
\label{Seff}
\end{equation}

\noindent
where ${\cal Z}_\psi$ is the wave-function renormalization function,
$\mu_3$ incorporates the mass renormalization function, ${\cal Z}_a$
incorporates the vertex renormalization function and $F_{\rm vacuum}$
represents the vacuum energy contributions coming from the integration
over the nonstatic Matsubara modes. The ${\cal O}\left[ \psi^*_0 \psi_0
|\nabla\psi_0|^2, (\psi_0^*\psi_0)^3\right]$ terms represent higher
order interactions in the zero modes of the fields. As shown in Ref.
\cite{second}, these terms will give contributions to the density of
order-$a^3$ and higher and therefore do not enter the order-$a^2$
calculations considered here. By matching perturbative order-$a^2$
results obtained with the original action $S_{E}$ and the general
effective action $S_{\rm eff}$, the authors of Ref. \cite{second} were
able to show that the transition temperature for a dilute, homogeneous,
three dimensional Bose gas can be expressed at next to leading order as

\begin{equation}
T_c= T_0 \left\{ 1 + c_1 a n^{1/3} + \left[ c_2^{\prime} \ln\left(a n^{1/3}
\right)+c_2^{\prime \prime} \right] a^2 n^{2/3} + {\cal O} \left(a^3 n
\right) \right\}\;,
\label {exptc}
\end{equation}
where $T_0$ is the ideal gas condensation temperature, $T_0 = 2 \pi/m
\left[n/\zeta(3/2)\right]^{2/3}$, $n$ is the density, $\zeta(x)$ is the
Riemann Zeta function and $c_1, c_2^{\prime}$ and $c_2^{\prime\prime}$
are numerical coefficients. A similar structure is also discussed in
Ref. \cite {markus}. As far the numerical coefficients are concerned,
the exact value, $c_2^{\prime}=-64 \pi \zeta(1/2) \zeta(3/2)^{-5/3}/3
\simeq 19.7518$, was obtained using perturbation theory \cite {second}.
On the other hand, the other two coefficients, $c_1$ and $c_2^{\prime
\prime}$, are sensitive to the infrared sector of the theory and
consequently cannot be obtained perturbatively, but they can, through
the matching calculation, be expressed in terms of the two
nonperturbative quantities $\kappa$ and ${\cal R}$ which are,
respectively, related to the number density $\langle \psi_0^* \psi_0
\rangle$ and to the critical chemical potential $\mu_c$, as shown below.
The actual relation between the two nonperturbative coefficients and
these physical quantities is given by \cite {second}

\begin{equation}
c_1 = - 128 \pi^3  [\zeta(3/2)]^{-4/3}  \kappa   \;\;,
\label{c1}
\end{equation}
and

\begin{equation}
c_2^{\prime \prime} = - \frac{2}{3} [\zeta (3/2)]^{-5/3} b_2^{\prime \prime} +
\frac {7}{9} [\zeta (3/2)]^{-8/3} (192 \pi^3 \kappa)^2 + \frac{64 \pi}{9}
\zeta (1/2) [\zeta(3/2)]^{-5/3}
\ln \zeta (3/2)   \;,
\label{c2}
\end{equation}
where $b_2^{\prime \prime}$, in Eq. (\ref{c2}), is

\begin{equation}
b_2^{\prime \prime} = 32 \pi \left \{ \left [ \frac{1}{2} \ln (128 \pi^3) +
\frac{1}{2} - 72 \pi^2 {\cal R} - 96 \pi^2 \kappa \right ]\zeta(1/2)
 + \frac {\sqrt {\pi}}{2} - K_2 - \frac {\ln 2}{2 \sqrt {\pi}}\left [
\zeta(1/2) \right ]^2 \right \}  \; ,
\label{b2primeprime}
\end{equation}
with $K_2= -0.13508335373$. The quantities $\kappa$ and ${\cal R}$ are
related to the zero Matsubara modes only. Therefore, they can be
nonperturbatively computed directly from the reduced action $S_{\rm
eff}$ which, as discussed in the numerous previous applications, can be
written as

\begin{equation}
S_{\phi}=  \int d^3x \left [ \frac {1}{2} | \nabla \phi |^2 +
\frac {1}{2} r_{\rm bare}
\phi^2 + \frac {u}{4!} (\phi^2)^2
\right ] \;,
\label{action2}
\end{equation}

\noindent
where $\phi = (\phi_1, \phi_2)$ is related to the original real
components of $\psi_0$ by $\psi_0({\bf x}) =\sqrt{mT/{\cal Z_\psi}}\;
[\phi_1({\bf x})+i\phi_2({\bf x})]$, $r_{\rm bare}=2m \mu_{3}/{\cal
Z_\psi}$ and $u=48 \pi a mT {\cal Z}_a/{\cal Z_\psi}^2$. The vacuum
contribution appearing in Eq. (\ref{Seff}) will not enter in the
specific calculation we do here and one has been omitted it from
Eq. (\ref{action2}). In the large-$N$ limit considered in
the first part of this work, and also in Refs. \cite {baymN,arnold}, the
field $\phi$ in Eq. (\ref{action2}) is formally considered as having $N$
components ($\phi_i$, $i=1,\ldots,N$). In this case, the Bose-Einstein
condensate effective action Eq. (\ref{action2}) is the $N=2$ special
case of the general $O(N)$ invariant action.

The three dimensional effective theory described by Eq. (\ref {action2})
is super-renormalizable and requires only a mass counterterm to
eliminate any ultraviolet divergence. In terms of Eq. (\ref{action2}),
the quantities $\kappa$ and ${\cal R}$ appearing in Eqs. (\ref{c1}) -
(\ref{b2primeprime}) are defined by

\begin{equation}
\kappa \equiv \frac {\Delta \langle \phi^2 \rangle_c}{u}  =
\frac { \langle \phi^2 \rangle_u - \langle \phi^2 \rangle_0}{u} \;,
\label{kappa}
\end{equation}
and

\begin{equation}
{\cal R} \equiv \frac {r_c}{u^2} = -\frac {\Sigma(0)}{u^2}  \;,
\label{R}
\end{equation}
where the subscripts $u$ and $0$ in Eq. (\ref{kappa}) mean that the
density is to be evaluated in the presence and in the absence of
interactions, respectively, and $\Sigma(0)$ is the self-energy with zero
external momentum. Since these physical quantities are dependent on the
zero modes their evaluation is valid, at the critical point, only when
done in a nonperturbative fashion. As discussed in the next section, the
relation between $r_c$ and $\Sigma(0)$ comes from the Hugenholtz-Pines
(HP) theorem at the critical point.

Eq. (\ref{exptc}) is a general order-$a^2$ result with coefficients
that, therefore, depend on nonperturbative physics via $\kappa$ and
${\cal R}$. In principle, to evaluate these two quantities one may start
from the effective three-dimensional theory, given by Eq. (\ref{action2}), 
and then employ any nonperturbative analytical or numerical
technique.

When quantum corrections are taken into account, the full propagator for the
effective three dimensional theory reads

\begin{equation}
G(p)=[p^2+ r +\Sigma_{\rm ren}(p)]^{-1} \;,
\end{equation}
where $p^2$ represents the three momentum and $\Sigma_{\rm ren}(p)$
represents the renormalized self-energies. At the transition point
($p^2=0$), the system must have infinite correlation length and one then
has

\begin{equation}
[G(0)]^{-1}=[r_c +\Sigma_{\rm ren}(0)] =0 \;.
\end{equation}
This requirement leads to the Hugenholtz-Pines theorem result
$r_c=-\Sigma_{\rm ren}(0)$. Since $r_c$ is at least of order $u$ it
would be treated as a vertex in a standard perturbation type of
calculation in which $G(p)=1/p^2$ represents the bare propagator. This
shows that perturbation theory is clearly inadequate to treat the BEC
problem at the transition due to the presence of infrared divergences.
One must then recur to nonperturbative methods like the numerical
lattice Monte Carlo simulations, the analytical $1/N$ or the linear
$\delta$ expansion (LDE) \cite {linear} adopted in this work (see for
instance Ref. \cite{early} for earlier works on the method).
The problem is highly nontrivial since the Hugenholtz-Pines theorem
automatically washes out all momentum independent contributions, such as
the one-loop tadpole diagrams, which constitute the leading order of
most approximations. In practice, this means that the first nontrivial
contributions start with two-loop momentum dependent self-energy terms.
However, having reduced the original model, Eq. (\ref{SE}) to the
effective three-dimensional one, Eq. (\ref{action2}), 
makes it easier to tackle those
contributions since one no longer has the problem of summing over the
Matsubara's frequencies, which is a hard task when the number of loops
increases.

Recent numerical Monte Carlo  
applications \cite{arnold1,russos} have predicted values for $c_1$
which are close to $1.30$ \footnote{See Ref. \cite{sun} for an extension 
of these
works to the $O(1)$ and $O(4)$ cases.}. On the other hand, some analytical
applications have predicted values such as $\sim 2.90$ obtained with a
self-consistent resummation method \cite {baymprl}, $\sim 2.33$ obtained
with the $1/N$ expansion to leading order  \cite {baymN} and
$\sim 1.71$ obtained with the same expansion to the next to leading
order \cite {arnold}. The LDE was first applied to
order-$\delta^2$ producing $c_1 \sim 3.06$ \cite {prb}. Recently, the
calculation has been extended to order-$\delta^4$ with the results $\sim
2.45$ at order-$\delta^3$ and $\sim 1.51$ at order-$\delta^4$ \cite
{pra}. The coefficient $c_2^{\prime \prime}$ was evaluated with Monte
Carlo techniques \cite {arnold1} and the predicted value obtained from
those simulations is $75.7\pm0.4$. This quantity was also analytically
evaluated with the LDE in Ref. \cite {pra}, where the encountered
numerical values are $\sim 101.4$, $\sim 98.2$ and $\sim 82.9$ at
second, third and fourth orders respectively. An order-$\delta^2$
application to ultra-relativistic gases has also been performed
\cite{tim}. The LDE has been especially successful in treating scalar
field theories at finite temperatures \cite{hatsuda,prd1} as well as
finite temperature and density \cite{hugh}. Several different
applications performed with the LDE are listed in Ref. \cite{pra}.
Recently, Braaten and Radescu \cite{braaten} have also used the LDE,
with different optimization prescriptions, to evaluate $T_c$ 
both at large and finite $N$ limits, while Kleinert \cite {kleinert} has
used the variational perturbation theory, which is a variation of the
LDE, obtaining the value $c_1 \sim 0.91 \pm 0.05$. Like we do here for 
the finite
$N$ case, he has considered up to order-$\delta^4$ contributions which
include five loop diagrams. However, none of those authors considers
resummation techniques to accelerate convergence and do not evaluate the
coefficient $c_2^{\prime \prime}$, that it is also computed in the 
present work.

As far as the application of the LDE to the determination of the BEC
transition temperature is concerned, since the
first papers making use of the LDE method to this problem
\cite{prb,pra}, an important question was raised and remained
unanswered. This question regards the convergence properties of the
method in this application for the BEC problem, which is in fact related
to the convergence of the method in critical theories in general.
Actually, this is a timely and important question regarding the
applications of the LDE in field theories, since the first efforts have
been concentrated mostly in the anharmonic oscillator  problem at
zero temperature, where rigorous LDE convergence proofs have been
produced \cite {phil,ian,kleinert2,guida}. The extension to the finite
temperature domain was also considered by Duncan and Jones \cite{dujo},
who used the anharmonic oscillator partition function. Only very
recently a convergence proof has been extended, but for a particular
perturbative series case, to asymptotically free renormalizable quantum
field theories at zero temperature \cite {jldamien}. 
Here, as in a
previous paper from us, Ref.
\cite {letter}, our interest is to probe convergence in the vicinity of
a phase transition, such as for the Bose-Einstein condensation problem
presented above. This study should also settle the questions regarding
the correctness of our original LDE applications \cite{prb,pra}, since
by showing convergence in the large-$N$ case we also establish the
reliability of our finite $N$ results, as originally studied in
those investigations. The LDE convergence in the large-$N$ extension of the
BEC problem has also been recently addressed by Braaten and Radescu \cite
{braaten}. The differences between their approach and ours will be
discussed in some detail in the text.

The literature shows \cite {phil,sunil,gastao} that in most applications
it is possible to establish simple relations between the LDE and other
nonperturbative methods already at order-$\delta$ where one-loop
diagrams are present. In fact, one can show that in those cases the LDE
either exactly reproduces $1/N$ results or produce very close numerical
results. Here, the BEC problem poses an additional difficulty since, as
discussed above, the first non-trivial contributions start at the
two-loop level in the self-energies. As we shall see, it is not easy in
this case to establish simple analytical relations for the quantities
being computed, like those given {\it e.g.} in
Refs. \cite{phil,sunil,gastao}, and the problem must be treated
differently. However, as we are going to show in the coming sections,
our numerical results, improved with an efficient resummation technique
exactly converge in the large-$N$ limit and seems to also converge in
the arbitrary $N$ case.

This paper is organized as follows. In Sec. II we briefly
recall the LDE method and present the
interpolated version of the action Eq. (\ref{action2}) to be studied
throughout the paper, following the recent applications performed in
Refs. \cite {prb,pra}. In Sec. III, we carry out the formal evaluation
of $\langle \phi^2 \rangle_u$ in three different ways. The first is the
usual order by order type of calculation, which is familiar from
perturbative calculations, and is in fact the only possible one for the
realistic finite $N$ case. The second uses the type of resummation
familiar from nonperturbative methods such as Hartree and the $1/N$
approximations. These two procedures generate two series for the
large-$N$ limit of $\langle \phi^2 \rangle_u$ (in which case $\phi$ in
Eq. (\ref{action2}) is extended to $N$ components) whose coefficients,
which are numerically obtained, can be usefully compared as a cross
check, helping to establish the numerical reliability of the finite $N$
series. In the same section, we also consider the asymptotic infrared
and ultraviolet behavior of the series which, as we shall see, is a very
useful approximation allowing at the same time a fully analytical
analysis. In Sec. IV we first examine the LDE optimized perturbation
procedure in a general case, considering a simple geometric series in
order to get insight regarding the convergence structure of the optimal
results. Then, the large-$N$ BEC series are optimized and a resummation
technique which accelerates convergence is introduced. By taking the
large-$N$ result for $c_1$ (obtained in Ref. \cite{baymN}) as a
reference value we proceed with the investigation of convergence,
showing that the LDE together with the
resummation technique can exactly reproduce the large-$N$ result already at
the first non-trivial order,
provided that it is applied to a specific approximation fully exploiting
the infrared limit properties. Having explicitly shown the LDE convergence
properties within a limit where an exact result exists we turn our
attention, in Sec. V, to the realistic finite $N$ case where no similar
infrared approximation is available. Using
the same resummation method, the order-$\delta^4$ results for $c_1$ and
$c_2^{\prime \prime}$ obtained in Ref. \cite{pra} are improved and the
results obtained seem to converge to the lattice Monte Carlo
estimates of Ref. \cite{arnold1}. Our conclusions are presented in Sec.
VI. {}For completeness and comparison purposes we also include two
appendixes, one where the original large-$N$ derivation is reviewed and
another one to detail useful properties of the large order behavior of
the LDE.

\section {LDE AND THE INTERPOLATED EFFECTIVE SCALAR THEORY FOR BEC}

Let us start our work by reviewing the application of the LDE method to
our problem. The LDE was conceived to treat nonperturbative physics
while staying within the familiar calculation framework provided by
perturbation theory. In practice, this can be achieved as follows.
Starting from an action $S$ one performs the following interpolation

\begin{equation}
S \rightarrow S_{\delta} =  \delta S + (1 - \delta) S_0(\eta) \;,
\label{s}
\end{equation}
where $S_0$ is the soluble quadratic action, added by an (optimizable)
mass term $\eta$, and $\delta$ is an arbitrary parameter. The
above modification of the original action somewhat
reminds the usual trick consisting of adding and subtracting a mass term
to the original action. One can readily see that at $\delta=1$ the
original theory is retrieved, so that $\delta$ actually works just as a 
bookkeeping parameter. 
The important modification is encoded in the field dependent
quadratic term $S_0(\eta)$ that, for dimensional reasons, must include
terms with mass dimensions ($\eta$). In principle, one is free to choose
these mass terms and within the Hartree approximation they are replaced
by a direct (or tadpole) type of self-energy before one performs any
calculation. In the LDE they are taken as being completely arbitrary
mass parameters, which will be fixed at the very end of a particular
evaluation by an optimization method. One then formally pretends that 
$\delta$ labels interactions
so that $S_0$ is absorbed in the propagator whereas $\delta S_0$ is
regarded as a quadratic interaction. So, one sees that the physical
essence of the method is the traditional dressing of the propagator to
be used in the evaluation of physical quantities very much like in the
Hartree case. What is different between the two methods is that, within
the LDE the propagator is completely arbitrary,  constrained only
to cope with the so-called direct terms ({\it i.e.} tadpoles) within the
Hartree approximation. So, within the latter approximation the relevant
contributions are selected according to their topology from the start.

Within the LDE one calculates in powers of $\delta$ as if it was a small
parameter. In this respect the LDE resembles the large-$N$ calculation
since both methods use a bookkeeping parameter which is not a physical
parameter like the original coupling constants and within each method
one performs the calculations formally working as if $N \rightarrow
\infty$ or $\delta \rightarrow 0$, respectively. {}Finally, in both
cases the bookkeeping parameters are set to their original values at the
end which, in our case, means $\delta=1$. However, quantities evaluated
at any finite LDE order from the dressed propagator will depend explicitly
on $\eta$, unless one could perform a calculation to all orders. Up to
this stage the results remain strictly perturbative and very similar to
the ones which would be obtained via a true perturbative calculation. It
is now that the freedom in fixing $\eta$ generates nonperturbative
results. Since $\eta$ does not belong to the original theory one may
require that a physical quantity $\Phi^{(k)}$ calculated perturbatively
to order-$\delta^k$ be evaluated at the point where it is less sensitive
to this parameter. This criterion, known as the Principle of Minimal
Sensitivity (PMS), translates into the variational relation \cite{pms}

\begin{equation}
\frac {d \Phi^{(k)}}{d \eta}\Big |_{\bar \eta, \delta=1} = 0 \;.
\label{pms}
\end{equation}

The optimum value $\bar \eta$ which satisfies Eq. (\ref{pms}) must be a
function of the original parameters including the couplings, which
generates the nonperturbative results. Another optimization procedure,
known as Fastest Apparent Convergence criterion (FAC) (see also Ref.
\cite{pms}), may also be employed. It requires, from the $k$-th
coefficient of the perturbative expansion

\begin{equation}
\Phi^{(k)} = \sum_{i=0}^k c_i \delta^i  \;,
\label{fac0}
\end{equation}
that

\begin{equation}
\left[\Phi^{(k)} - \Phi^{(k-1)}\right]\Bigr|_{\delta=1} =0\;,
\label{fac}
\end{equation}
which is just equivalent to taking the $k$-th coefficient
(at $\delta=1$) in Eq. (\ref{fac0}) equal to zero.
{}For the interested reader Refs.
\cite{prb,pra,tim,hatsuda,prd1,hugh,ian,kleinert2,guida,phil,dujo,jldamien,letter,sunil,gastao,braaten}
provide an extensive (but far from complete) list of successful 
applications of the method to different problems.

It is important to recall that the basic reason for the convergence of the
LDE method in the quantum mechanics case (anharmonic oscillator
energy levels typically) \cite{phil,ian,kleinert2,guida} relies on the 
fact that 
the LDE modifies perturbative expansions in such a way that
the PMS or FAC optimized values of the initially arbitrary 
mass parameter (the equivalent of $\eta$ here) essentially follow, at large
perturbative orders,  a pattern of rescaling this mass with
the perturbative order, which is such as to compensate
the generic factorial growth \cite{ao} of the original perturbative expansion
coefficients at large orders. As we will see in Sec III, in the
present BEC case the relevant perturbative expansions do not exhibit
factorially growing coefficients, but nevertheless the reasons for convergence
of the LDE share some similarities with the latter cases, since the LDE
followed by PMS also introduces at sufficiently large order
a certain scaling behavior with the
perturbative order, in such a way as to modify
(extend) the convergence radius of the original perturbative series. 

Let us now write the interpolated version of the effective model
described by Eq. (\ref{action2}). Before doing that let us rewrite
$r_{\rm bare} = r + A$ where $A$ is a mass counterterm coefficient
needed to remove the UV divergence from the self-energy. This
counterterm is the only one effectively needed within the modified
minimal subtraction ({$\overline {\rm MS}$}) renormalization scheme
adopted here for the evaluation of the relevant {}Feynman diagrams
contributing to the self-energy, as explicitly shown in Ref. \cite{pra}.
Then, one can choose

\begin{equation}
S_0 =  \frac{1}{2}  \left [ | \nabla \phi|^2 + \eta^2 \phi^2  \right ] \;,
\label{S0}
\end{equation}
obtaining

\begin{equation}
S_{\delta}=  \int d^3x \left [ \frac {1}{2} | \nabla \phi |^2 +
\frac {1}{2} \eta^2 \phi^2  +\frac{\delta}{2}(r - \eta^2) \phi^2 +
\frac {\delta u}{4!} (\phi^2)^2   + \frac {\delta}{2} A_{\delta} \phi^2
\right ] \;\;,
\label{Sdelta}
\end{equation}

\noindent 
where $A_{\delta}$ represents the renormalization mass counterterm
for the interpolated
theory, which depends on the parameters $\eta$ and $\delta$. It is
important to note that by introducing only extra mass terms in the
original theory the LDE does not alter the polynomial structure and,
hence, the renormalizability of a quantum field theory. In practice, the
original counterterms change in an almost trivial way so as to absorb
the new $\eta$ and $\delta$ dependence. The compatibility of the LDE
with the renormalization program has been shown in the framework of the 
$O(N)$ scalar
field theory at finite temperatures, in the first work of Ref. \cite{prd1}, 
showing that it consistently takes into account anomalous
dimensions in the critical regime. Note also that we have treated $r$
as an interaction, since this quantity has a
critical value ($r_c$) that is at least of order $\delta$.

Requiring the original system to exhibit infinite correlation length at
the critical temperature, means that, at $T_c$ and $\delta= 1$ (the
original theory), the full propagator $G^{(\delta)} (p)$, given by

\begin{equation}
G^{(\delta)}(p)=
\left[ p^2 + \eta^2 + \delta r -
\delta{\eta^2} + \Sigma^{(\delta)}_{\rm ren}(p) \right  ]^{-1}\;,
\label {G}
\end{equation}
must satisfy $G^{(\delta)}(0)^{-1} =0$. This requirement implies

\begin{equation}
\delta r^{(\delta)}_c = -\Sigma^{(\delta)}_{\rm ren}(0)\;,
\label{HP}
\end{equation}
which is equivalent to the Hugenholtz-Pines theorem
applied to the LDE.

\section{LDE EVALUATION OF $\langle \phi^2 \rangle_u^{(\delta)}$ 
IN THE LARGE-$N$ LIMIT}

Let us now turn our attention to the explicit LDE evaluation of $\langle
\phi^2 \rangle_u^{(\delta)}$ in the large-$N$ limit. In practice, the
large-$N$ evaluation can be performed in different fashions which
include the conventional order by order perturbative evaluation and the
more economical closed form evaluation in which the whole large-$N$
series is resummed. The first, purely perturbative method in the
standard Feynman graph way is also the only possible one concerning the
finite $N$ calculations, where different classes of diagrams contribute.
The second technique is usually employed in approximations such as
Hartree and $1/N$ where it is possible to sum a certain class of graphs
based on the type of loop terms they contain. Having resummed a given
class, one may easily obtain a perturbative result by expanding the
series to a given order in $\delta u$. Of course, both methods must lead
to equivalent analytical results but, as we shall see, the final
numerical results can be different at high perturbative orders. This is
due to the fact that both perturbative expansions contain coefficients
numerically produced. Since our optimization procedures may be
sensitive to the numerical precision of those coefficients it will be
instructive to compare them in detail. {}Finally, one can move one step
further by obtaining a series with exact coefficients that allows for a
fully analytical investigation. This is made possible by considering an
approximation which avoids complicated integrals appearing in the exact
calculation due to the presence of dressed propagators in terms of
self-energies, which are usually cumbersome beyond some given order.
Such a simpler series, with exact coefficients, is obtained if one
considers typically the physically motivated deep infrared behavior of
the dressed scalar propagators. In this section we explore these three
possible evaluations.

\subsection{Standard Perturbative Evaluation of 
$\langle \phi^2 \rangle_u^{(\delta)}$}

Let us evaluate $\langle \phi^2 \rangle_u^{(\delta)}$ in the usual
perturbative way. The relevant contributions, in the large-$N$ limit,
are shown in {}Fig. 1. Using the full propagator one may write this
quantity, at the critical point, as

\begin{equation}
\langle \phi^2 \rangle_u^{ (\delta)}= \sum_{i= 1}^{N}
\langle \phi^2_i \rangle_u^{ (\delta)} =  N
\int \frac {d^3 p}{(2 \pi)^3} G^{(\delta)} (p) =
\int \frac {d^3 p}{(2 \pi)^3}
\frac{N}{p^2 +(\eta^{*})^2}\left [ 1  + \frac {\delta r^{(\delta)}_c  +
\Sigma^{(\delta)}_{\rm ren}(p) }{p^2 +(\eta^{*})^2} \right ]^{-1} \;,
\label{p2}
\end{equation}
where $\eta^* = \eta \sqrt{1-\delta}$. Note that with this prescription
one only has to evaluate diagrams which would appear in a usual
perturbative calculation since the quadratic $\delta \eta^2$ vertex is
automatically taken into account when $\eta^*$ is expanded to the
relevant order in $\delta$.

One can express the large-$N$ calculation more conveniently, in the
generalization of
Eq. (\ref{action2}) to $O(N)$ symmetry, by
considering $u= u^{\prime}/N$. In this case the nontrivial contributions, 
in the large-$N$ limit and expanded to LDE order $k$, are given by

\begin{equation}
\langle \phi^2 \rangle_u^{(k)} = N \int \frac {d^3 p}{(2 \pi)^3}
\frac{1}{p^2 +(\eta^*)^2}
- N \int \frac {d^3 p}{(2 \pi)^3}
\frac {[\Sigma^{(n)}(p)- \Sigma^{(n)}(0)]}{[p^2 +(\eta^*)^2]^2}   \;,
\label{phi2u}
\end{equation}
where we have used Eq. (\ref{HP}) and $\Sigma^{(n)}$ denotes
the $n$-bubble self-energy given by

\begin{equation}
\Sigma^{(n)}(p)=-\frac{2}{N} \left(-\frac{\delta u^{\prime}}{6}\right)^{n+1}
\int \frac {d^3 l}{(2 \pi)^3}\frac {1}{(l^2 +(\eta^{*})^2)}
\left [
\int \frac {d^3 s}{(2 \pi)^3}\frac {1}{(s^2 +(\eta^{*})^2)} 
\frac {1}{[(s+p-l)^2 +(\eta^{*})^2]} \right ]^n   \;,
\label{sigman}
\end{equation}
which is then of order $k=n+1$ in $\delta$.
Note that the mass counterterm is
a redundant quantity in the evaluation of $\langle \phi^2
\rangle_u^{(k)}$ because this quantity depends on the difference

\begin{equation}
\Sigma^{(n)}_{\rm ren}(p)-\Sigma^{(n)}_{\rm ren}(0)  =
[ \Sigma^{(n)}_{\rm div}(p)+ \Sigma^{(n)}_{\rm ct}(p)]-
[\Sigma^{(n)}_{\rm div}(0)  + \Sigma^{(n)}_{\rm ct}(0) ] \;,
\end{equation}
where $\Sigma^{(n)}_{\rm div}(p)$ is the divergent self-energy. {}For a
general renormalizable theory, the quantity $\Sigma^{(n)}_{\rm ct}(p)$
represents all counterterms associated with the parameters of the theory
(such as masses and coupling constants) as well as the wave-function
counterterm associated with any eventual momentum dependent pole. At the
same time, $\Sigma^{(n)}_{\rm ct}(0)$ involves the same counterterms
except for the wave-function one. However, as we have already emphasized
previously, in the three-dimensional case the only type of primitive
divergence requires only a mass counterterm, which is the same for
$\Sigma^{(n)}_{\rm div}(p)$ and $\Sigma^{(n)}_{\rm div}(0)$. This means
that in our case, $\Sigma^{(n)}_{\rm div}(p)-\Sigma^{(n)}_{\rm div}(0)$
is always a finite quantity. It turns out that this quantity is also
scale independent as discussed in Ref. \cite{pra}. Since however the
individual contribution $\Sigma^{(n)}_{\rm div}(p)$ contain a
divergence, we regularize all diagrams with dimensional regularization
in arbitrary dimensions $d= 3-2\epsilon$, where in the modified minimal
subtraction ($\overline{\rm MS}$) renormalization scheme, 
the momentum integrals can be written as

\begin{equation}
\int \frac {d^3 p}{(2 \pi)^3} \to
 \left(\frac{e^{\gamma_E} M^2}{4 \pi} \right)^\epsilon
\int \frac {d^d p}{(2 \pi)^d} \;,
\label{rdim}
\end{equation}

\noindent
where $M$ is an arbitrary mass scale and $\gamma_E \simeq 0.5772$ is the
Euler-Mascheroni constant.

Then, from the use of standard  {}Feynman parameters for the integrals
over momenta, we can
write the general form for  each of the two terms in  $\langle \phi^2
\rangle_u^{(k)}$, Eq. (\ref{phi2u}), that depend on the self-energy.
The first of such term can be expressed as

\begin{eqnarray}
-  N \int \frac {d^3 p}{(2 \pi)^3} 
\frac {\Sigma^{(n)}(p)}{[p^2 +(\eta^*)^2]^2}  
&=&
\frac {(- \delta u^{\prime})^{n+1}} {3(6\eta^*)^n}  \frac
 {\Gamma[n(1/2+\epsilon)+2\epsilon ]}{(4\pi)^{3/2(n+2)}}
\left ( \frac {e^{\gamma} M^2}{(\eta^*)^2} \right )^{\epsilon(n+2)} 
\nonumber  \\
&\times& \int_0^1 d \chi \chi(1-\chi)^{-3/2+\epsilon+n(1/2+\epsilon)} 
\int_0^1 d \gamma (1-\gamma)^{n(1/2+\epsilon)-1}
[\gamma(1-\gamma)]^{1/2-\epsilon - n(1/2+\epsilon)} \nonumber \\
&\times& \int_0^1 d\alpha_1 [\alpha_1(1-\alpha_1)]^{-(1/2+\epsilon)} \ldots
\int_0^1 d\alpha_n [\alpha_n(1-\alpha_n)]^{-(1/2+\epsilon)}
\nonumber \\
&\times&\int_0^1 d\beta_1 \beta_{1}^{n-2} [\beta_1(1-\beta_1)]^{-1/2} 
\int_0^1 d\beta_2 \beta_{2}^{n-3}
[\beta_1\beta_2(1-\beta_2)]^{-1/2} \ldots \nonumber \\
&\times& \int_0^1 d\beta_{n-1}
[\beta_1\beta_2 \ldots
\beta_{n-1}(1-\beta_{n-1})]^{-1/2} 
{\cal F}(\chi,\gamma,\alpha_i,\beta_j) \;,
\label{phip}
\end{eqnarray}
where

\begin{equation}
{\cal F}(\chi,\gamma,\alpha_i,\beta_j)=
\left \{ \chi + \frac{(1-\chi)}{\gamma(1-\gamma)}
\left [ \gamma+(1-\gamma)
\left ( \frac {(1-\beta_1)}{\alpha_1(1-\alpha_1)} 
+ \frac {\beta_1(1-\beta_2)}{\alpha_2(1-\alpha_2)} + \ldots +
\frac {(\beta_1\beta_2 \ldots \beta_{n-1})}{\alpha_n(1-\alpha_n)} \right ) 
\right ] \right \}^{-[n(1/2+\epsilon)+2\epsilon]}
\;.
\end{equation}
At the same time, the $p=0$ term is given by

\begin{eqnarray}
- N \int \frac {d^3 p}{(2 \pi)^3} 
\frac {\Sigma^{(n)}(0)}{[p^2 +(\eta^*)^2]^2}  
&=&
\frac {(- \delta u^{\prime})^{n+1}} {3(6\eta^*)^n}
\frac {\Gamma[n(1/2+\epsilon)+\epsilon -1/2]}{8\pi(4\pi)^{3/2(n+1)} }
\left ( \frac {e^{\gamma} M^2}{(\eta^*)^2} \right )^{\epsilon(n+1)} 
\nonumber  \\
&\times&  \int_0^1 d \gamma (1-\gamma)^{n(1/2+\epsilon)-1}
[\gamma(1-\gamma)]^{1/2-\epsilon - n(1/2+\epsilon)} \nonumber \\
&\times& \int_0^1 d\alpha_1 [\alpha_1(1-\alpha_1)]^{-(1/2+\epsilon)}
\ldots \int_0^1 d\alpha_n [\alpha_n(1-\alpha_n)]^{-(1/2+\epsilon)}
\nonumber \\
&\times&\int_0^1 d\beta_1 \beta_{1}^{n-2} [\beta_1(1-\beta_1)]^{-1/2} 
\int_0^1 d\beta_2 \beta_{2}^{n-3}
[\beta_1\beta_2(1-\beta_2)]^{-1/2} \ldots \nonumber \\
&\times& \int_0^1 d\beta_{n-1}
[\beta_1\beta_2 \ldots  \beta_{n-1}(1-\beta_{n-1})]^{-1/2} 
{\cal G}(\gamma,\alpha_i,\beta_j) \;,
\label{phi0}
\end{eqnarray}
where

\begin{equation}
{\cal G}(\gamma,\alpha_i,\beta_j)=\left \{  \frac{1}{\gamma(1-\gamma)}
\left [ \gamma+(1-\gamma)
\left ( \frac {(1-\beta_1)}{\alpha_1(1-\alpha_1)} + 
\frac {\beta_1(1-\beta_2)}{\alpha_2(1-\alpha_2)} + \ldots +
\frac {(\beta_1\beta_2 \ldots \beta_{n-1})}{\alpha_n(1-\alpha_n)} \right )
\right ] \right \}^{-[n(1/2+\epsilon)+\epsilon -1/2]}
\;.
\end{equation}
It is not very difficult to see, by counting the superficial degree of
divergence in Eq. (\ref{sigman}), that the only ultraviolet divergence
shows up in the one-bubble ($n=1$) contribution. In Eq. (\ref {phip}) the UV
divergence for this case hides in the term
$\chi(1-\chi)^{-3/2+\epsilon+n(1/2+\epsilon)}$ and appears explicitly
upon integration by parts over $\chi$. After that one can take the usual
expansion in powers of $\epsilon$ and perform a numerical integration
over the {}Feynman parameters to obtain for the first, non-vanishing,
term in Eq. (\ref{phip}) the result \cite {prb}

\begin{equation}
-N\int_{p}\frac{\delta^2 \Sigma^{(1)}(p) }{[p^2 + (\eta^*)^2]^2}=
\delta^2\frac{  (u^{\prime})^2}{\eta^*} \frac{1}{18(8\pi)^3}
\left [  \frac{1}{ \epsilon}+6\ln \left (\frac {M}{2\eta^*}
\right ) + 2 - 4 \ln 2 \right ]\;.
\label{ssp}
\end{equation}
In the $p=0$ case the pole shows up in the gamma function which 
becomes $\Gamma(2\epsilon)$ for $n=1$. Integration yields

\begin{equation}
N\int_p\frac {\delta^2 \Sigma^{(1)}(0)}{[p^2 +(\eta^*)^2]^2}=
- \delta^2 \frac{(u^{\prime})^2}{ \eta^*}\frac{1}{18 (8\pi)^3}
\left [ \frac{1}{\epsilon}  + 6 \ln \left (\frac {M}{2 \eta^*} \right ) 
+2 + 4 \ln(2/3)\right ]
\;.
\label{ss0}
\end{equation}
The last two equations also reproduce the results found analytically in
Refs. \cite {prb,brani}. As already mentioned, although Eq.
(\ref{ssp}) and Eq. (\ref{ss0}) diverge, their sum is finite and
scale independent. Together, they give the contribution

\begin{equation}
-N\int_{p}\frac{\delta^2 [\Sigma^{(1)}(p)-\Sigma^{(1)}(0)]}{[p^2 + (\eta^*)^2]^2}=
-{\delta^2} \frac {(u^{\prime})^2}{\eta^*} \frac {1}{18(8\pi)^3} 4 \ln(4/3)\;.
\label{braaten}
\end{equation}
All higher loop contributions are finite and one can safely take
$\epsilon=0$ in Eqs. (\ref {phip}) and (\ref{phi0}). Note that in the
above perturbative series, that is generated order by order from the
standard Feynman graph procedure, the first non-trivial contributions
start at order $\delta^2 u^2$, due to the fact that the first order
expansion term, linear in $\delta u$ ({\it i.e.} $n=0$ in Eq. (\ref{sigman})),
is automatically canceled as a consequence of the Hugenholtz-Pines
theorem, Eq. (\ref{HP}). The number of
{}Feynman variables at each order is $k+2$ for Eq. (\ref {phip}) and
$k+1$ for Eq. (\ref {phi0}). We then get the order-$\delta^{20}$ result
in the large-$N$ limit

\begin{equation}
\langle \phi^2 \rangle_u^{(20)}= - \frac {N \eta^*}{4\pi} + 
\delta \frac {u N}{3}
\sum_{i=1}^{19} C_i \left ( - \frac {\delta u N
}{6 \eta^*}
\right )^i  + {\cal O} (\delta^{21}) \;.
\label{exp20}
\end{equation}
Except for the first coefficient, where we have the exact result from 
Eq. (\ref{braaten}), all the other coefficients for $i \ge 2$ can only
be obtained numerically. One well known numerical routine that can be 
used to evaluate the $i$-dimensional integrals over the {}Feynman parameters
in  Eqs. (\ref {phip}) and (\ref{phi0}) is the
Monte Carlo multidimensional
integration routine VEGAS \cite{vegas}. However, one should bear in mind that
VEGAS may not be so reliable for a very large number of dimensions,
since VEGAS, as a Monte-Carlo integration method, 
inherently makes use of finite numbers of points and iterations
and these cannot be increased indefinitely in practice
in order to improve precision. So
for integrals with a {\em very} large number of dimensions (for example the
last coefficient $C_{19}$ involve a 39-dimensional integral) the VEGAS may
lead to wrong estimates for both the numerical value of the integral and the
corresponding error bar estimate obtained from the code (which also depends on the number
of points and iterations used). {}Fortunately, as we will see below,  all the terms
contributing at large $N$
can be computed alternatively in a much easier way (and to
arbitrary precision) in terms of
one-dimensional integrals, thus assuring a much better precision
for the results and we do not need to worry about any specific detail
of any numerical routine to evaluate  Eqs. (\ref {phip}) and (\ref{phi0}).
The coefficients obtained this way, which we denote by
$J_i$ and are given in the following Subsection, will be the results
used in all of our large-$N$ calculations. 
These $J_i$ coefficients can be contrasted then to the results obtained
from e.g. with VEGAS, that we show here for illustrative purposes only,
obtained using $10^4$ points with $100$ iterations fixed VEGAS parameters:
$C_1=(7.249 \pm 0.001)\times10^{-5}$, 
$C_2=(2.050 \pm 0.003)\times 10^{-6}$,
$C_3=(6.32 \pm 0.01)\times 10^{-8}$,
$C_4=(2.048 \pm 0.003)\times 10^{-9}$,
$C_5=(6.85 \pm 0.01)\times 10^{-11}$, 
$C_6 = (1.709 \pm 0.002)\times 10^{-12}$, 
$C_7 = (3.561 \pm 0.006) \times 10^{-14} $, 
$C_8= (6.48 \pm 0.01) \times 10^{-16} $, 
$C_9= (1.063 \pm 0.002) \times 10^{-17}$,
$C_{10}= (1.560 \pm 0.005) \times 10^{-19}$, 
$C_{11}= (2.59 \pm 0.04) \times 10^{-21} $, 
$C_{12}= (5.00 \pm 0.09)\times 10^{-23}$, 
$C_{13}= (9  \pm 2)\times 10^{-26}$,
$C_{14}= (5 \pm 1) \times 10^{-28}$, 
$C_{15}= (5 \pm 1) \times 10^{-30}$,
$C_{16}= (1 \pm 7) \times 10^{-33} $, 
$C_{17}= (2.3  \pm 0.6)\times 10^{-34}$, 
$C_{18}= (2 \pm 2) \times 10^{-37} $ and 
$C_{19}= (5 \pm 5 ) \times 10^{-39} $.
Note that starting with $C_{13}$ the errors increase considerably as the 
dimension increases, as expected with a fixed
number of VEGAS parameters.

\subsection{Closed Form Evaluation of $\langle \phi^2 \rangle_u^{(\delta)}$}

Let us now write the whole large-$N$ perturbative series in a closed
form which resembles the usual $1/N$ resummation of {}Feynman graphs.
{}For completeness and comparison purposes we re-derive, in  Appendix A,
the original large-$N$ result found by Baym, Blaizot and Zinn-Justin
\cite {baymN}. Note, as already emphasized, that one basic difference
between the original large-$N$ calculation and the LDE one is that the
latter automatically introduces an infrared regulated propagator, from
the explicit mass term $\eta$. Apart from its main purpose of defining
in that way the relevant LDE series in $\delta u/\eta$, cf. Eq.
(\ref{exp20}), this also has the advantage of explicitly regularizing
the intrinsic infrared divergence of the corresponding expression of
$T_c$ in the original calculations \cite{baymN,arnold}. However, similarly
with the latter, there still remain some subtleties with this
closed (resummed) form of the perturbation series, related to the fact
that the integrals over momenta are not absolutely (UV) convergent, as
we shall examine below. Thus, after applying the Hugenholtz-Pines
theorem, and summing all the leading large-$N$ contributions shown in
{}Fig. 1, one obtains for the expression equivalent to Eq. (\ref{phi2u})
the result

\begin{eqnarray}
\langle  \phi^2 \rangle^{(\delta)}_u &= & N \int \frac {d^3 p}{(2 \pi)^3}
\frac{1}{p^2 +(\eta^*)^2}  \nonumber \\
&-& \frac {\delta u N}{3}
\int \frac{d^3\,p}{(2\pi)^3} \;
\frac{d^3\,k}{(2\pi)^3} \;
\frac{1}{[{p}^2+(\eta^*)^2]^2}\;
\left [1 + \frac { \delta u N}{6} B({k},\eta^*) \right ]^{-1}\;
\left[\frac{1}{({k} + {p})^2+(\eta^*)^2}-
\frac{1}{{k}^2+(\eta^*)^2 }\right ]   \;,
\label{basicN}
\end{eqnarray}
where
\begin{equation}
B(k,\eta^*) = \int \;\frac{d^3
q}{(2\pi)^3}\;\frac{1}
{[q^2+(\eta^*)^2]\;[({k} + {q})^2+(\eta^*)^2]}
\;\; = \frac{1}{4\pi k}\;
 \arctan \left( \frac{k}{2\eta^*}\right )  \;,
\label{sigex}
\end{equation}
with $k \equiv |{\bf k}|$ and similarly for $p$, $q$ in Euclidean space. 
Contrary to the corresponding expression in the massless case $\eta =0$
(see Appendix A), here it
appears not possible to integrate exactly Eq. (\ref{basicN}) due to the
non-trivial dependence in $k$ and $\eta^*$ of the resummed propagator
$B(k,\eta^*)$. But one can at least still do exactly the first integral over
the momentum $p$. In Eq. (\ref{basicN}), the integral over $p$ is finite in
$d=3$, and can be easily performed to give

\begin{equation}
 \langle \phi^2 \rangle^{(\delta)}_u = -\frac {N \eta^*}{4\pi} -  
\frac {\delta u N}{3}
\frac{1}{(8\pi \eta^*)}
\int \frac{d^3 k}{(2\pi)^3}
\left[ 1 +\frac {\delta u N}{6} B({k},\eta^*)\right ]^{-1}
\; \left [\frac{1}{{k}^2+4(\eta^*)^2}-\frac{1}{{k}^2+(\eta^*)^2 }\right]\;,
\label{basic2N}
\end{equation}
while the remaining $k$-integral can be performed numerically. 
After a little algebra one gets

\begin{equation}
\langle \phi^2 \rangle_u^{(\delta)}= -\frac {N\eta^*}{4\pi} + \frac {u N}
{96\pi^2} + \delta \frac {u N}{3}
\sum_{i=1}^{\infty} J_i \left ( - \frac {\delta u N
}{6 \eta^*}
\right )^i   \;,
\label{exp20spurious}
\end{equation}
where the $J_i$ coefficients are given by

\begin{equation}
J_i=  \frac {3}{16 \pi^3} \left ( \frac {1}{8\pi} \right )^i
\int_0^{\infty} dz \frac {z^2}{(z^2+1)(z^2+4)}[A(z)]^i  \;,
\label{Js}
\end{equation}
with

\begin{equation}
A(z) = \frac {2}{z} \arctan \frac {z}{2} \;,
\end{equation}
and $z=k/\eta^*$. Analytically, the two ways we have presented for
obtaining the perturbative evaluation of $\langle \phi^2
\rangle_u^{(\delta)}$ are expected to be equivalent and any difference
may only arise from the numerical evaluation of the $C_i$ and $J_i$
coefficients. In this respect one expects $J_i$, which are evaluated
from one dimensional integrals, to be more accurate than $C_i$ and it
will be instructive to compare both results. We have numerically
evaluated $J_i$ with both MATHEMATICA \cite{matha} and Maple, where
we can compute the integrals with arbitrary precision, using
diverse integration routines available in both, in order to check the
reliability of the results. The first nineteen values obtained are
$J_1=7.24858\times 10^{-5}$, $J_2=2.04919 \times 10^{-6}$,
$J_3=6.32139\times 10^{-8}$, $J_4=2.04829 \times 10^{-9}$, $J_5=6.85295
\times 10^{-11}$, $J_6 = 2.3454 \times 10^{-12}$, $J_7 =8.16524 \times
10^{-14}$, $J_8=2.88069 \times 10^{-15}$, $J_9= 1.02726 \times
10^{-16}$, $J_{10}= 3.69589 \times 10^{-18}$, $J_{11}= 1.33955 \times
10^{-19}$, $J_{12}= 4.88611 \times 10^{-21}$, $J_{13}= 1.79203 \times
10^{-22}$, $J_{14}= 6.60406 \times 10^{-24}$, $J_{15}= 2.44405 \times
10^{-25}$, $J_{16}= 9.07903 \times 10^{-27}$, $J_{17}= 3.38400 \times
10^{-28}$, $J_{18}= 1.26514 \times 10^{-29}$ and $J_{19}= 4.74288 \times
10^{-31}$. Comparing $C_i$ to $J_i$ one sees that the multi-dimensional
VEGAS routine produces accurate results up to $i = 5$ but the values
quickly deteriorate at large orders. This is a consequence that the
VEGAS routine does not handle well integrals with a too large number of
dimensions, for relatively (and computationally viable) small number of
points and iterations, as explained at the end of Sec. III.A. 
Therefore, in this work, we shall consider only the series
with the more accurate $J_i$ coefficients, as computed from Eq.
(\ref{Js}). 

However, one notes that Eq.~(\ref{exp20spurious}) displays a more
significant difference with respect to the ``direct" perturbative series
in Eq.~(\ref{exp20}), which is due to the presence of the extra first
order term in $\delta u$, independent of $\eta$. Remark that this
contribution is just the {\em opposite}, in sign, of the exact large-$N$
result \cite{baymN} (see Eq. (\ref{exactN}) in Appendix A). This
apparent difference between the two approaches to the large-$N$
perturbative series deserves a detailed discussion to which we now turn
our attention on. In fact, the above difference is only a consequence of
integrating expression (\ref{basicN}) over $p$ first. Namely, if
performing the expansion in powers of $\delta$, and then integrating
first over $k$ in Eq.~(\ref{basicN}) (which is formally equivalent to
what is done in the standard perturbative order-by-order graphical
approach in the previous subsection), the linear term in $\delta u$
automatically cancels out. More precisely one obtains in this case an
integral of the type ($\xi$ just denotes an arbitrary mass parameter
here, that is equal to $\eta^*$ in the LDE calculation)

\begin{equation}
\sim N\,u \delta \; \int \frac{d^3\,k}{(2\pi)^3} \;
\left(\frac{1}{({k} + {p})^2+\xi^2}-
\frac{1}{{k}^2+\xi^2 }\right ) \;,
\label{otherint}
\end{equation}
and in dimensional regularization the two terms in the bracket just
cancel out, as can be seen by making a shift $k \to k-p$ in the first
term. (One can also check with a standard cutoff regularization that the
integral in Eq. (\ref{otherint}) gives a zero result, though it is a
less immediate calculation). Therefore depending on the order in which the two
integrals in Eq.~(\ref{basicN}) are performed one may get different
results, which is precisely the manifestation of an ambiguity due to the
fact that the integrals are not absolutely (UV) convergent, as pointed
out in Refs. \cite{baymN,arnold}. Actually, this problem is more basically
rooted in the fact that in obtaining Eq. (\ref{basicN}) one has formally
resummed a series containing UV divergences, considering, e.g., the
separate contributions in the last parenthesis of Eqs. (\ref{basicN})
(see also Eqs. (\ref{basm0}) and (\ref{sigma}) in Appendix A).
Therefore, the actual point is that one is not allowed (in principle) to
exchange the perturbative, all order summation, with integration, which
in our calculation is reflected in the different resulting perturbative
series. Going back to the standard perturbative expansion, as performed
in the previous subsection (Eqs.~(\ref{phi2u})--(\ref{phi0})) the
perturbative parameter expansion in powers of $\delta$ is made first,
and the UV divergence (which only appears at first non-trivial order
$\delta^2 u^2$ as discussed there) may be taken care of by the standard
renormalization. On the other hand, if we formally re-expand Eq.
(\ref{basicN}) in a power series in $u$, we can immediately see that the
integral defining the coefficient of the first order term linear in $u
\delta$, originating from $[1+(\delta u N/6)B(k,\eta^*)]^{-1} \sim 1 $,
has momenta routing that cannot be consistent with the actual
perturbative graph: rather, considering for instance the first term of
the last parenthesis in Eq. (\ref{basicN}), one should have $p+k \to p$
for consistency (see {}Fig.~\ref{fig1}), since this first order term in
$\delta u$ implies that the resummed propagator (dotted line) is pinched
to a point, so that there is no $k$-momentum flow. This would give again
a zero result for the coefficient of $u$, just for the same reasons as
Eq.~(\ref{otherint}) is vanishing, while formally performing the
integral with $p+k$ instead gives the $N\,u/(96\pi^2)$
term~\footnote{Remark that in the original calculation\cite{baymN}
corresponding to take $\eta =0$ in Eq. (\ref{basicN}), this ambiguity
problem has been consistently solved simply by using dimensional
regularization (see Appendix A for details): in contrast, the rather
subtle point is that when $\eta \neq 0$ in Eq. (\ref{basicN}) it
appears at first perfectly consistent to start with the $p$ integration.
Nevertheless, this does not give the correct result, independently of
whether one uses dimensional regularization or not, until one correctly
identifies what the actual perturbative series in $\delta u/\eta$
should be, as explained above.}. To summarize, while it will appear very
convenient to formally perform the integral over $p$ first in
Eq.(\ref{basicN}), it is only consistent provided one subtracts the
spurious linear term from the naive result, Eq. (\ref{exp20spurious}).
The correct perturbative series thus reads 

\begin{equation}
\langle \phi^2 \rangle_u^{(\delta)}= -\frac {N\eta^*}{4\pi}
+ \delta \frac {u N}{3}
\sum_{i=1}^{\infty} J_i \left ( - \frac {\delta u N
}{6 \eta^*}
\right )^i   \;,
\label{exp20braaten}
\end{equation}
which has the same form as   Eq. (\ref{exp20}).  
An expression similar to Eq. (\ref{exp20braaten}) was
also found by Braaten and Radescu in Ref. \cite{braaten}.

\subsection{Asymptotic infrared and ultraviolet behavior of $ \langle \phi^2
\rangle^{(\delta)}_u$}

Before considering the relevant BEC perturbation series,
Eq.~(\ref{exp20}), or equivalently Eq. (\ref{exp20braaten}), in the
large-$N$ limit, let us recall some expected general properties of the
large order perturbative expansions, as seen from a diagrammatic point
of view. This digression will emphasize an important difference between
the generally expected large order behavior of perturbative series in
most renormalizable models and the behavior of the above BEC specific
series. 

In field theory one has to face the problem of the perturbative series
being often only an asymptotic (non-convergent) series which, in most
cases, is due to factorial growing perturbative coefficients at large
orders. If the coefficients are of the same sign, order by order, those
series are not even Borel summable \cite{Borel,renormalons}. In
practice, this means that the perturbative expansion alone does not
define uniquely the physical quantities being expanded, so that the
series has to be complemented by intrinsically nonperturbative
contributions, containing typically terms with an exponential dependence
in the (inverse) expansion parameter \cite{renormalons}. This is
problematic because, apart from the special case of exactly
solvable/integrable models, in most theories those nonperturbative terms
are at best known only on phenomenological grounds. However, to
investigate the large order behavior of the perturbative series (and
therefore guess at least the form of nonperturbative missing
contributions), it is often sufficient to consider a class of
approximated graphs, expected (and proved in some specific models) to
give the dominant contributions to the perturbative coefficients at
large order. In dimension $d=2,3$ and $4$ renormalizable theories, such
dominant graphs are typically given by the next-to-leading term in a
$1/N$ expansion, where, roughly speaking, the matter fields are in a
$N$-vector representation, and it is sufficient to consider the
asymptotic behavior in the bubble ($1/N$) approximation of the relevant
Green functions. {}For instance, in a theory with a renormalized
coupling one obtains, after renormalization, a scale-dependent running
coupling. A very sketchy Green's function calculation in the above
approximation involves (after renormalization) typical momentum
integrals of the form

\begin{equation}
\int dq F(q^2) g(M) \left [ 1 - \beta_0 g(M) \ln
\frac{q^2}{M^2}\right ]^{-1}  \;,
\label{basren}
\end{equation}
where $F(q^2)$ is model-dependent and characteristic of the
Green's function considered, $g(M)$ is the running
coupling, $\beta_0 $ is the first order renormalization group (RG)
$\beta$-function coefficient:
$d\,g(M)/d\ln M \equiv \beta_0 g^2(M)+\cdots$,
and $M$ is an arbitrary renormalization scale. When formally expanding
Eq.~(\ref{basren})  in a perturbative series in $g$,  one gets
integrals of the form

\begin{equation}
\sum_p g^{(p+1)}(M) (\beta_0)^p
\int_0^\infty d q^2 F(q^2) \ln^p \left (\frac{q^2}{M^2} \right )   \;,
\label{intp}
\end{equation}
which leads to a factorial behavior $p!$ at large $p$. More precisely,
$F(q^2)$ can be expanded in a power series in $q^2$ ($1/q^2$) in the
infrared (respectively in the ultraviolet), so that Eq.~(\ref{intp})
gives series of the form \cite{renormalons} $\sim
g^{(p+1)}(-\beta_0)^p\; p !$ or $\sim g^{(p+1)}(\beta_0)^p\; p !$
for large $p$, respectively, for the infrared and ultraviolet
behavior. Considering for example an
asymptotically free theory, {\it i.e.} with $\beta_0 <0$, one obtains a
non sign-alternated series thus non-Borel summable, as far as the
infrared behavior is concerned. This fact reflects the important
infrared sensitivity of such theories.

Now, a drastic difference between the previous illustration of theories
with a renormalized coupling and the effective BEC $\phi^4$ model in
three dimensions considered here, is that for the latter only the mass
is renormalized, so that the coupling is finite and dimensionful, as
pointed out previously. {}From this, and following the
above line of reasoning, one expects that the relevant BEC $T_c$
perturbative series coefficients at large orders do not display any
factorial behavior. Therefore, a more convergent series should appear,
as is confirmed for instance by the form of the exact large-$N$
perturbative series in Eqs. (\ref{exp20}), (\ref{exp20braaten}), whose
coefficients appear clearly not very different from those of a geometric
series. Also, from the above general considerations, a similar behavior
of the series is expected as well for arbitrary $N$. Thus an interesting
question is whether one could obtain from such large order behavior
estimates a sensible approximation of the exact series in $\delta u/\eta$,
that would be relevant within the LDE method.

Let us therefore investigate some analytically simpler but physically
motivated approximations (expected to be asymptotically dominant) of the
large order behavior of the power series in $\delta u/\eta$, as generated
from Eq.~(\ref{basic2N}). $B({k},\eta)$ behaves as

\begin{equation}
B({k},\eta^*) \sim
\left\{
\begin{array}{ll}
\frac{1}{8\pi}\;\frac{1}{\eta^*}(1-\frac{k^2}{12(\eta^*)^2}+\cdots) ,
\;\; & (k \ll \eta^*)
\\
& \\
\frac{1}{8\pi}\;(\frac{\pi}{k} -\frac{4\eta^*}{k^2}+\cdots),
\;\; & (k \gg \eta^*)
\end{array}
\right.
\label{sigir-uv}
\end{equation}
for  the IR ($k \ll \eta^*$) and
UV ($k \gg \eta^*$) limits, respectively. This means that the
deep IR behavior of the series should be essentially given by
a $k$-independent term, $[1+\delta u N/(48\pi\eta^*)]^{-1}$, 
replacing the corresponding bracket in Eq. (\ref{basicN}). 
Retaining only this simplest IR behavior, the remaining
integral over $k$ becomes straightforward and accordingly 
we obtain\footnote{Note that, in close analogy with Eqs. (\ref{basren}),
(\ref{intp}), we only approximate the non-trivial resummed propagator
$B(k, \eta^*)$ according to Eq. (\ref{sigir-uv}), and keep the
exact $k, \eta^*$ dependence of the remaining integrand.}  
the  relevant IR approximated series as a simple geometric
series

\begin{equation}
\langle  \phi^2 \rangle_{IR}^{(\delta)} = -\frac {N\eta^*}{4\pi}
- \frac{\delta u N}{24\pi\eta^*}
\left [1+\frac{\delta u N}{48\pi\eta^*}\right ]^{-1}\:\left (-\frac{2\eta^*}{4\pi}
+\frac{\eta^*}{4\pi}\right ) -\frac{\delta u N }{96\pi^2}
= -\frac {N\eta^*}{4\pi} +\frac{\delta u N }{96\pi^2}\;\left
[\left (1+\frac{\delta u N}{48\pi\eta^*}\right  )^{-1}\;-1 \right] \;, 
\label{asymN}
\end{equation}
where the first order term, independent of $\eta$, has been explicitly
subtracted from the naive integral result Eq.~(\ref{basic2N}) for
consistency, as discussed in detail in the previous Subsec. III.B (note
that the purpose of the last parenthesis in the first equality in Eq.
(\ref{asymN}) is to retain, for clarity, the separate contributions of
the two propagator terms in Eq. (\ref{basic2N}) ). Note also that Eq.
(\ref{basic2N}) is UV finite, so that the result Eq. (\ref{asymN}) is
independent of the integration method used, so that either dimensional
regularization or another integration method leads to the same result.

Similarly, we can still integrate Eq. (\ref{basicN}) exactly when taking
the UV limit of the propagator, $B(k,\eta^*) \sim (8 k)^{-1}$ from
Eq.~(\ref{sigir-uv}). One obtains

\begin{equation}
\displaystyle
\langle  \phi^2 \rangle_{UV}^{(\delta)} = -\frac {N\eta^*}{4\pi} +
\left (\frac{N \eta^*}{2\pi^3}\right )\:
\left[ \pi \,y
\left(7 +4\,y^2\right)
 -8\,\ln (2)\left(1+
y^2\right)- 6\,\ln y \right]\;
     \left( 1 + 5\,y^2
  +4\,y^4
        \right)^{-1} \;-\frac{\delta u N }{(96\pi^2)}\;,
\label{asymUV}
\end{equation}
where $y \equiv 48\,\eta^*/(N\,\delta\,u)$. {}When performing the $p$
integration first, all integrals are UV finite in $d=3$ and again the $\eta
\neq 0$ mass explicitly regularizes the IR divergences. Note for instance that
in both, Eqs.~(\ref{asymN}) and (\ref{asymUV}), the $1/\eta^*$ and $\eta^*$
from the first and second integrals respectively, cancel out: this is
completely analogous to the cancellation in the original calculation
\cite{baymN}, in dimensional regularization with no IR cutoff mass, of the
pole $1/(d-3)$ (see Appendix A for details). 

Next, remark that by taking the $\eta\to 0$  limit of expression
(\ref{asymN}) or (\ref{asymUV}), one recovers the correct $1/N$ exact
result Eq.~(\ref{exactN}). In other words, we see here that the massless
limit $\eta\to 0$ of the LDE series consistently reproduces the exact
large-$N$ result, which is expected since the LDE for $\delta \neq 1$
plays the role of an infrared regulator. But this check is important as
regards the question of the possible convergence of the LDE series
to the exact result, once a non-trivial optimization of the LDE
series, with respect to the arbitrary remaining mass parameter $\eta$,
will be performed, to be examined in the next section. 

With this aim, it is instructive to re-expand, in a power series of
$\delta u/\eta^*$, the above two different IR and UV approximations for
$\Delta \langle \phi^2 \rangle^{(\delta)} $. {}First, taking the IR
limit Eq. (\ref{asymN}) gives the geometric series in $\delta u/\eta^*$:

\begin{equation}
\langle  \phi^2 \rangle_{IR}^{(k)} = -\frac {N\eta^*}{4\pi}
+ \frac{ \delta u \; N }{3} \sum_{i=1}^k G_i
\left (- \frac { \delta u \, N}{6\eta^*} \right )^i
\label{serIR}
\end{equation}
where $G_i \equiv [(64\pi^2)(8\pi)^i]^{-1}$. Numerically, the first five $G_i$
coefficients are $G_1= 6.2991\times 10^{-5}$, $G_2 = 2.5063 \times
10^{-6}$, $G_3= 9.9724 \times 10^{-8}$, $G_4 = 3.9679 \times 10^{-9}$
and $G_5=1.5788 \times 10^{-10}$, which are interesting to compare with
the corresponding exact coefficients in Eq.~(\ref{exp20braaten}). One
can see that the first low order coefficients of Eqs. (\ref{serIR}) and
(\ref{exp20braaten}) are of very similar magnitudes, and we have further
checked that significant departures ({\it i.e.} about an order of
magnitude or more) between the IR-approximated and exact large-$N$
perturbative coefficients only occur at rather large, greater than
${\cal O}(\delta^{15})$, orders. In other words, one expects from this
comparison the IR-approximated series, which has a convenient and
simpler geometric form, to be a very good approximation of the exact
large-$N$ series. This is a strong indication that the detailed
non-asymptotic (infrared) behavior of the scalar propagator should play
essentially no role, as could be physically expected on general grounds,
and as will be fully confirmed by our numerical investigation below. 

The same expansion for the UV limit (\ref{asymUV}) reads similarly 

\begin{equation}
 \langle  \phi^2 \rangle_{UV}^{(\delta)} = -\frac {N\eta^*}{4\pi}
 +\frac{ \delta \:u N}{3} \left [ -1.7465 \times 10^{-4} \left (
\frac{\delta u}{3\eta^*} \right ) + 2.4737 \times 10^{-5} 
\left (\frac{\delta u}{3\eta^*}
\right )^2 +\cdots \right ] \;.
\label{serUV}
\end{equation}
where one can see in contrast that the coefficients are already quite
different from the exact series Eq.~(\ref{exp20braaten}) at low order,
so that we may expect the asymptotic UV limit of the propagator to give
a less sensible approximation than the IR one. {}For completeness it is
useful to consider alternatively the direct evaluation of
Eq.~(\ref{basicN}) ({\it i.e.} not as a perturbation series in $\delta
u/\eta$). Taking thus the exact expression for $B(k,\eta^*)$ instead
of its simpler IR or UV limits, the $k$-integration can be performed
only numerically (note that it is still IR and UV finite in $d=3$).
Those (numerical) results for the reference value $u=1$ and as function
of $\eta^*$ are illustrated in Fig. \ref{fig2}.  As one can
see, both the IR and UV approximations, as given in Eq.~(\ref{sigir-uv}),
have a  behavior that is somewhat
different from the exact function Eq.~(\ref{basicN}) 
for very small $\eta^*$, although all
expressions correctly give the exact result at $\eta^* =0$. However,
though it is not visible on Fig.2, the IR and UV approximations 
appear to be very good approximations of the exact function for 
larger $\eta^* \sim {\cal O}(1)$, that is in the range where their respective
perturbative expansion forms start to be valid.

To summarize this subsection, introducing an IR regulator mass $\eta
\neq 0$ as is done from the LDE procedure, together with the deep IR
limit of the propagator, Eq. (\ref{sigir-uv}) in Eq.~(\ref{basicN})
leads to perturbative series that are very close to the exact large-$N$
one in (\ref{exp20braaten}). The only subtlety when implementing the LDE
within the convenient resummed large-$N$ closed form Eq.~(\ref{basicN}),
is to remind that one should be careful in exchanging the perturbative
series summation with integration, since the resulting integral, Eq.
(\ref{basicN}), is not (absolutely) UV convergent. The consequence is
that, {\it e.g.}, the $\eta^*\to 0$ limit of expression (\ref{basic2N})
does not commute with taking the limit $\eta^*\to 0$ {\em before}
performing any integration, {\it i.e.} as is done in the original
large-$N$ calculation (see Appendix A). This may be considered as a
reminiscence of the infrared sensitivity of the theory, even if it is
not as severe as the above mentioned models with a running coupling,
leading to divergent series with factorial growing perturbative
coefficients.

\section{ LARGE-$N$ OPTIMIZATION, RESUMMATION AND CONVERGENCE PROPERTIES}

As discussed in the introduction, the study of LDE convergence properties in
the BEC problem is much more complicated than in the pure anharmonic
oscillator case \cite{ian,kleinert2,guida,phil}. In principle, both
models are described by a scalar $\phi^4$ model in one and three
dimensions but in the BEC case, the model is used to study a phase
transition. However, if the LDE works, one expects that reasonable
numerical results should be obtained, converging to the ``exact"
large-$N$ result, $c_1=8 \pi /[3 \zeta(3/2)^{4/3}]\simeq 2.328$,
evaluated in Ref. \cite{baymN}. Before proceeding, it is useful to point
out an essential aspect and potential difficulty of the purely
perturbative expansions, Eq. (\ref{exp20braaten}), or similarly Eq.
(\ref{serIR}). Note that these are {\em inverse} (alternated) series in
the mass parameter $\eta$, thus with an a priory finite convergence
radius $ \eta_c^{-1}$ (i.e. for $u=1$ these series are (absolutely)
convergent only for $|\eta| >\eta_c$) when considered as series for
complex values of the arbitrary mass parameter $\eta$. But, as discussed
above, ultimately the exact result is expected to be recovered for $\eta
\to 0$. This situation is not much different from the anharmonic
oscillator case, where typically the energy levels have perturbative
expansions in powers of $\lambda/m^3$ ($\lambda$ being the coupling and
$m$ the mass) with moreover factorially
growing coefficients at large perturbative orders, but where nevertheless the
LDE converges \cite{phil,ian,kleinert2,guida} to the exact result, thanks to
an appropriate rescaling of the mass parameter that is consistent
with the PMS optimized solutions. We will
examine here how the LDE procedure followed by the standard PMS
optimization, Eq. (\ref{pms}), manage in fact to avoid this $\eta\to 0$
potential problem with the basic perturbation series, which is one of
the main results of the present paper. This is where the infrared
approximation is a useful guide: while its perturbative form Eq.
(\ref{serIR}) exhibits just the same feature as the exact series, Eq.
(\ref{exp20braaten}), the former geometric series is known to all
orders, and obviously its $\eta \to 0$ limit is perfectly well-defined
and gives the correct exact result, as can be seen by going back to its
original form Eq. (\ref{asymN}) discussed in Sec. III.C. But no such
resummation is a priory known for the relevant non-trivial $N=2$ series,
where only the first few perturbative orders in $u/\eta$ are known and
thus only the latter information can be used to define the LDE
procedure. 

Thus, before considering the LDE of the actual BEC series, it appears
very instructive to first examine the same LDE procedure performed on a
simpler model which shares many similarities with the relevant BEC
problem. We will see that the example below illustrates very well the
basic reasons for the success (or eventually failure) of the LDE
followed by the PMS optimization method in the general case, beyond the
specific BEC problem considered in this paper. 

\subsection{A simple example of the LDE-PMS convergence}

Let us examine the properties of the LDE and subsequent PMS optimization
in a general case by considering the following function
which admits a simple alternated
geometric series expansion:

\begin{equation}
\Phi(x)=  -\frac{1}{x} -\frac{x}{(1+x)}\;= -\frac{1}{x} +\sum^\infty_{n=1}
(-x)^n  
\label{simple}
\end{equation}
which expansion form is almost like our IR geometric series
Eq.~(\ref{serIR}), for $u=1$
and $x \equiv 1/\eta^*$, apart from overall different normalizations
(compare {\it e.g.} with Eq. (\ref{exp20braaten}) or (\ref{serIR})).
Clearly, the exact expression in Eq.~(\ref{simple}) tends to $-1$
for $x\to \infty$, thus the goal is to examine whether the LDE procedure
followed by the standard PMS optimization, which only uses the series
expansion form in Eq.~(\ref{simple}) 
at successive perturbative orders, is able to reach such
a result and in which way. To make contact with the LDE series we
consider the series in Eq. (\ref{simple}) with $x \equiv u\delta/\eta^*$
(except in the $-1/x$ term on the right-hand side of Eq.~(\ref{simple})
where we
take $x \equiv 1/\eta^*$), followed by an expansion in power series of
$\delta$ in which one takes $\delta\to 1$. The result of this LDE at arbitrary
order $k$ can be expressed entirely analytically in this case (see also
Appendix B)\footnote{Note a slight difference in Eq. (\ref{simplde}) with
respect to Eq. (\ref{ldep}) in Appendix B, namely 
the Gamma function ratio $\Gamma[-n/2+k]/\Gamma[-n+k]$ in place of
$ \Gamma[1-n/2+k]/\Gamma[1-n+k]$ appearing in Eq.~(\ref{ldep}), 
due to the fact that
Eq.~(\ref{simple}) does not have the extra factor of $u\delta$ in front
of the series. Because of this, non-trivial PMS solutions
$d\Phi(x)^{(k)}/dx =0$ of Eq.~(\ref{simplde}) start already at LDE order
$k=1$, see Table 1. But apart from that, this difference only affects in
very minor ways the qualitative behavior discussed below, in particular
the large LDE order behavior of the simple series (\ref{simplde}) with
respect to the corresponding actual BEC series.}:

\begin{equation}
\Phi(x \equiv u/\eta)^{(k)}= -\frac{1}{x} \sqrt{\pi}
\frac{(-1)^k}{\Gamma[1/2-k]\Gamma[1+k]} + \sum^k_{n=1}
(-x)^n\:  \frac{\Gamma[1-n/2+k]}{\Gamma[1-n+k]\Gamma[1+n/2]}
\label{simplde}
\end{equation}
and
the PMS optimization performed order by order (up to order $k=10$)
is shown in Table I. From this table, we can draw
several important observations:\\

\begin{description}

\item i) Applying the PMS optimization condition Eq. (\ref{pms}) to
$\Phi(x)$ generates complex $\bar \eta$ solutions. In addition, all
solutions can be arranged into families which span the complex plane. In
general, a new family arises at even orders and this is signaled by a
first member which lies in the real axis. This pattern was also found in
the anharmonic oscillator applications \cite {phil} and in the finite
$N$ applications to the BEC case \cite {pra}.

\item ii) Despite the fact that the (naive) expansion at arbitrary
finite order has a finite convergence radius $|x| < 1$ (namely, the LDE
at order $k$ only uses the information of the right-hand side of Eq.
(\ref{simple}), where the series is absolutely convergent for $|x|<1$),
the LDE-PMS procedure clearly converges to the right result for
$x\to\infty$ (equivalently $\eta \to 0$) and the convergence is quite
rapid in this case.

\item iii) Though most of the PMS solutions do converge to the correct 
$x\to\infty$ result, there is a family of real solutions, $S0$, that clearly
converges  to a different result ($\sim 2$). 

\item iv) Some of the (complex) solutions converge more rapidly than
others. In particular, at order 3 one of the solutions is already very
close to the exact result, and also one of the two real solutions that
does converge to the exact result, which appears only at odd LDE orders,
is not the best one. Note also that it is the real part which converges
to the right result. However, to obtain those results one must
consider all, real and complex, $\bar{ \eta}$ values. In fact, the
imaginary parts of the complex optimized values $\bar \eta$ play an
essential role and for instance suppressing even a small imaginary part
results in a completely wrong and unstable result for the optimized
series. As we shall investigate in more detail below, this is because,
for convergence of the series, it is the value of $|x|$ (respectively
$|\eta|^{-1}$) which is relevant. 

\end{description}

We shall now investigate in more details the basic reasons for the main
LDE convergence results ii) above. First, note that in Eq.
(\ref{simple}) we added an extra term $1/x \sim \eta^*$, since it
vanishes for $x\to\infty$. In the relevant BEC case, as already
discussed in the previous Section, this term has a very clear physical
interpretation, as it corresponds to the tadpole graph. In the present
case it may be considered only as a mathematical trick, which plays an
important role for convergence. Indeed, when suppressing this extra
linear ``mass" term, one finds for the LDE-PMS in place of the results
shown in Table I: $-0.5625$ and $-0.6574 \pm 0.153 I$ {\it e.g.} at LDE
orders $k=2$ and $k=3$ respectively; and at order $k=10$ the best
solution is $-0.97208 \pm 0.285 I$ while other solutions are still about
$20\%$ away from the exact result. Thus, the additional linear term
clearly increases substantially the speed of convergence. First, it is
clear that non-trivial PMS solutions already start at order 1, while
they start only at order 2 when suppressing this term. But this is not
the main reason for this faster convergence. Remark that at very large
order, say $k \sim 100$, the numerical accuracy from both procedures
({\it i.e.} with or without the additional linear term) tend to become
essentially equivalent (except that the incorrect solution $\sim 2$, see
iii) above, is absent in the procedure with the linear term suppressed).
However, it is evidently crucial to have a procedure such that the very
first few LDE orders give already reasonably reliable results, and also
to gain an order in the PMS solution, when we will consider the relevant
$N=2$ BEC series for which only the first few perturbative orders are
known. The essential role of the linear term is easily understood when
expanding the (exact) result in Eq. (\ref{simple}) for large $x$: 

\begin{equation}
\Phi(x)= -\frac{1}{x} -\frac{x}{(1+x)} = -\frac{1}{x} -\left (1+\frac{1}{x}
\right )^{-1} =-1+ {\cal O}(1/x^2)
\end{equation}
such that for $x\to\infty$ ($\eta\to 0$) the first order ${\cal O}(1/x)$
term cancels out. Now what happens is that this cancellation remains
after introducing the LDE procedure, thus leading to a faster LDE-PMS
convergence, see remarks ii) above, even though the LDE modifies the
structure of the perturbative expansion and, as already mentioned, only
uses the perturbative information from Eq.~(\ref{simple}). This can be
understood on basic grounds when considering the large $k$ behavior of
expression (\ref{simplde}): 

\begin{equation}
\Phi(x \equiv u/\eta)^{(k)} \raisebox{-0.4cm}{~\shortstack{ 
$\sim$ \\ $k\to\infty $}}
 -\frac{1}{x\:k^{1/2}\,\Gamma[1/2]}
+ \sum^k_{n=1} \frac{(-x)^n\:k^{n/2}}{\Gamma[1+n/2]}\;\;
\sim -\frac{1}{x\:k^{1/2}\Gamma[1/2]} +\exp(x^2 k)\:Erfc(x k^{1/2}) -1
\label{simpldeinf}
\end{equation}
where $Erfc(x)$ is the standard error function \cite{abramowitz}, and
this large order behavior in (\ref{simpldeinf}) is obtained after some
algebra by using standard properties of the Gamma functions (see also
Appendix B). Furthermore, it appears that this large $k$ behavior of Eq.
(\ref{simpldeinf}) exhibiting the scaling $k^{1/2}$ is rapidly
reached: for instance the difference between
$(-1)^k/\Gamma[1/2-k]/\Gamma[k+1]$ in Eq.~(\ref{simplde})
and $ k^{-1/2}/\pi$ in Eq.~(\ref{simpldeinf}) is already $<
4\%$ for $k \ge 3$. Analyzing thus from
Eq.~(\ref{simpldeinf}) the large $k$ behavior,
one
can derive by using the asymptotic expansion of $Erfc(x)$ for the relevant
limit $x\to\infty$: $\exp{(x^2)}\:Erfc(x) \sim 1/(\sqrt{\pi}x)\:[1+{\cal
O}(1/x^2)]$ (see also Eq.~(\ref{Erfas}) in Appendix B), that the first order
term ${\cal O} (1/x/k^{1/2})$ compensates exactly the linear term, giving in
Eq. (\ref{simpldeinf}) $\Phi(x )\sim -1 +{\cal O}(1/x^2/k)$, while omitting the
linear term gives instead: $\Phi(x )\sim -1 +{\cal O}(1/x/k^{1/2})$.
Though the latter properties
result from taking the somewhat extreme limit $k\to\infty$, what is
quite remarkable is that this behavior is already well observed at very
low orders, as the above comparison of numerical PMS optimization in
Table I illustrates. 
 
More generally, one can also understand from Eq. (\ref{simpldeinf})
the main transformation operated
by the LDE on the original series: while the latter had a finite convergence
radius, for $|\eta| > 1$, the LDE provides an extra damping factor
$1/\Gamma[1+n/2]$ which allows to safely reach larger $|x|$ values
(equivalently smaller $|\eta|$ values) from the new perturbative series, so
that eventually the relevant limit $\eta\to 0$ may be approached. 
Indeed in practice, namely at finite $k$ orders, all of the PMS
solutions $\bar\eta$ of Eq.~(\ref{pms}) corresponding to Table I
(except the ones in the first column corresponding to the incorrect solution,
as will be discussed below)
tend to have smaller and smaller $|\bar \eta|$ values (though rather slowly
decreasing) as $k$ is increased.  

A last remark on the LDE behavior of the simple example (\ref{simple})
concerns the occurrence of the incorrect PMS solution, as indicated in
iii) above. Clearly, this results directly from the presence of the
additional linear term: in absence of the latter, these extra solutions
disappear from the LDE-PMS at any orders. Actually, these are
reminiscent from the fact that Eq.~(\ref{simple}) (before LDE is
performed), has another extremum at $x =-1/2$ ({\it i.e.} $\eta =-2$).
In the most general case where one would have no a priory idea {\it
e.g.} on the sign of the correct solution, this feature may be
considered as a drawback of our procedure. But in fact, it is easy to
get rid of this incorrect solution, simply because one knows that the
solution we seek should be for $\eta\to 0$. In contrast, the
PMS optimized values at successive LDE orders $k$
corresponding to this extra 
solution are always such that $ |\bar\eta|$ is
maximal, with respect to all other solutions. Furthermore, these $
|\bar\eta|$ values do not exhibit the expected trend towards
smaller and smaller values: on the
contrary, the corresponding $|\eta|$ value is (rather slowly) increasing 
as $k$ is increased. 

The important point is that, as we will see below, all of the above
properties will be exhibited similarly by the more complicated BEC
series and are, therefore, a very useful guideline. In particular, the
cancellation due to the additional linear term as above observed, the
behavior with the LDE order $k$ of the PMS solutions, including the behavior
of the incorrect PMS solution, all occur similarly in the more complicated
cases of the actual BEC LDE series, with an expected much faster LDE
convergence. These are important remarks concerning the LDE applications
since, as already emphasized, the difference between our original
applications Refs. \cite{prb,pra} and the one performed in Ref.
\cite{braaten} amounts to the fact that the latter authors optimize
$\Delta \langle \phi^2 \rangle^{(\delta)}$ 
in Eq.~(\ref{kappa}) which differs from $\langle
\phi^2 \rangle^{(\delta)}$ by an equivalent linear term.

In the next subsections, we switch to the study of the LDE performed on
the actual BEC series in the large-$N$ limit. With this purpose we will
consider $ \langle \phi^2 \rangle^{(\delta)}_u$ as given by the closed
form series, Eq. (\ref{exp20braaten}), and alternatively also its
infrared approximation form, Eq. (\ref{serIR}). These two quantities
will be again optimized with the standard PMS criterion, Eq.
(\ref{pms}). {}For completeness we also show, with one example, how the
alternative {}Fastest Apparent Convergence (FAC) method, Eq.
(\ref{fac}), generates similar results. We then present in Sec. IV.C a way to
further rearrange the LDE series, by fully exploiting the above
explained behavior of the PMS optimized solutions, which
eventually accelerates further the convergence due to the fact that 
it can recover
directly the large LDE order behavior and consequent good $\eta \to 0$
properties of the infrared approximation, with the advantage of being
applicable in a general case to more arbitrary series. This technique is
then used to treat both the complete closed form and the infrared
approximation of $ \langle \phi^2 \rangle_u^{(\delta)}$, as introduced
respectively in Secs. III.B and III.C. For comparison purposes an alternative 
resummation
procedure, based on Pad\'e approximants \cite{pade}, is also studied in Sec.
IV.D. {}Finally, we analyze all results to conclude in Sec IV.E about the
convergence structure of  the LDE method.

\subsection {Standard optimization}

In order to perform a numerical analysis of the optimized solutions for
$\langle \phi^2 \rangle_u^{(k)}$ one first expands Eq. (\ref {exp20}) to
the desired order in $\delta$. As usual, one then sets $\delta=1$ before
optimizing. This is the LDE part of the procedure. It can be performed
either by explicit order by order expansion, or equivalently more
formally following the general structure of the LDE expansion at
arbitrary order $k$ as presented in Appendix B, leading to the
result Eq.~(\ref{ldep}). Like the simple geometric series considered
above, Eq. (\ref{simple}), the equation for $\langle \phi^2
\rangle_u^{(k)}$ is basically a series in powers of $(\delta
u)^k/\eta^{(k-1)}$, so that the PMS and Fastest Apparent Convergence
procedures will also generate algebraic equations of order $k$ whose
mathematically acceptable roots form a set of optima $\bar \eta$ values.

Applying the PMS condition, Eq. (\ref{pms}), 
to $\langle \phi^2 \rangle_u^{(k)}$,

\begin{equation}
\frac {d \langle \phi^2 \rangle_u^{(k)}}{d \eta}\Big |_{\bar \eta, \delta=1} 
= 0 \;,
\label{pms1}
\end{equation}
or the Fastest Apparent Convergence criterion, $\langle \phi^2
\rangle_u^{(k)} - \langle \phi^2 \rangle_u^{(k-1)} =0$, computed at
$\delta=1$ (which is again analogous of taking the $k$-th coefficient in
Eq. (\ref{fac0}), with $\Phi^{(k)} = \langle \phi^2 \rangle_u^{(k)}$,
equal to zero) generates the optimal results. Table II lists, to
order-$\delta^{20}$ all families obtained by applying the PMS
optimization to the standard closed form expansion given by Eq. (\ref
{exp20braaten}). {}For completeness we compare, in Table III, results
generated with the PMS optimization and the {}Fastest Apparent
Convergence procedures for the same quantity. This table clearly shows
that both optimization criteria generate compatible solutions that seem
to converge to identical results. We advance that this pattern was also
observed in all remaining cases considered in this work, so we shall
consider only the PMS optimization, Eq. (\ref{pms1}), from now on. In
addition, we can get more insight by considering our infrared
approximated geometric series, Eq. (\ref {serIR}), which has simple
exact coefficients known to all orders. 
By applying the standard PMS, Eq. (\ref{pms1}), to
this simpler equation one obtains the results shown in Table IV. This
table has exactly the same characteristics as Table II except for the
positive real part numerical values, that are much closer to the exact
large-$N$ value $c_1 \simeq 2.32847$. 

Let us try to examine at this stage the (eventual) convergence structure
of the LDE as applied to the large-$N$ limit series, namely the results
presented in Tables II-III and IV. In both cases the family structure is
very similar to the one found for the simple geometric series example in
the previous subsection, as well as in the studies that have proven the
LDE convergence in the anharmonic oscillator \cite {phil}. For instance,
we also see in the present case the appearance of a new family at even
orders, such that all of them start with a real solution and become
complex at the next order, the exception being $F0$ which has only
negative real solutions and $F4$ which has only positive real solutions.
As mentioned before, the complex parts arise as a consequence of solving
the polynomial equations generated during the optimization procedure and
are mathematically acceptable. We refrain from trying to attach any
physical significance to the complex part of the optima $c_1$ values and
instead of considering only completely real solutions we take the
optimized ${\rm Re}(c_1)$ as the relevant quantity for the evaluation of
$T_c$, which is a strictly real quantity. Also, in the geometric series
application above, we saw that only the real parts of the optimal
solutions converge to the expected real value when $x \to \infty$.
Considering all complex solutions has also the advantage of giving a
prediction at any order. {}Further, we note that all families whose real
parts are positive start with values close to $\sim 2.0$ and then seem
to follow similar patterns as the perturbative order increases. In
contrast, considering $F0$ would bring important qualitative changes
regarding the critical temperature shift in relation to the ideal gas
value, $\Delta T_c=T_c-T_0$, since the sign of this quantity has also
been a source of controversy \cite {markus} for some time until
recently, when most works started to predict positive values for the
critical temperature shift. It appears clearly that $F0$ is just the
equivalent of the $S0$ (wrong) solution discussed in the simplest
geometric series case, which was eliminated because it does not
correspond to the $\eta\to 0$ expected behavior. As we will see below,
the same criteria allows to eliminate this solution without ambiguity.
Pushing further to much higher orders we obtain for the exact $1/N$
series case Eq. (\ref {exp20braaten}), e.g. at LDE order 100, that most
of the positive roots give a solution whose (real part) is $\sim
2.2$, as well as a few solutions which are far apart reasonable
values, which we assume only reflect the numerical limitation of the
problem at hand. Indeed, to obtain those results the numerical procedure
involves first to numerically solve the one dimensional integrals
related to the $J_i$ coefficients in Eq. (\ref {exp20braaten}). As
emphasized, this has been done with great care with the maximum
precision allowed by MATHEMATICA \cite {matha} and/or Maple. Yet, one
cannot expect the results to be completely stable once the LDE is
carried out to very high orders. Though the integration in
Eq.~(\ref{Js}) may be in principle
done to arbitrary accuracy, the limitation comes about later in the
process of finding the roots of high order polynomial equations, which
is a notoriously unstable numerical problem in general. So it is rather
the optimization procedure, as implied by the LDE/PMS which is to be
carried out numerically, which appears 
sensitive to numerical accuracy at very high orders.
Nevertheless from our analysis all results performed with MATHEMATICA
appear under very good control, say until a LDE order of about $\sim 50$. 
It
looks that all (stable) families in Table II will (slowly) converge, at
very high orders, to approximately the same values, in a way again
similar to what was observed in the simple geometric series example
above discussed in Sec. IV.A and in the anharmonic oscillator case
\cite {phil}. However, it appears that these PMS optimized solutions in
Table II rather converge to a value of about $\sim 2.2$, {\it i.e.}
close to but not equal to the exact large-$N$ result $c_1 \simeq
2.328$. The reason for this slight discrepancy will be understood
below. 

In contrast, the results in Table IV illustrate how good the IR
approximation is, although from a perturbatively inequivalent approach,
and indicates that the LDE series in Eq. (\ref{asymN}) do converge
to the correct result. Pushing the LDE to higher orders, we obtain
for Eq.~(\ref{serIR}), for instance at order 100, that most of the
solutions are very close (within $0.1 \%$ error) to the exact result,
the closest solution being $2.32834 \pm 0.00132884 I $.\\

In order to explain the convergence properties on more basic grounds,
both for the exact $1/N$ or IR approximated series, one first observes
that, similarly to the simple example studied in Sec IV.A above, all of the
PMS optimization solutions corresponding to Tables II-IV, except for $F0$,
have $|\bar \eta|$ 
starting from relatively large values and then (rather slowly)
decreasing as the LDE order $k$ is increased,  therefore reaching the boarder
of the convergence radius of the original $\delta u/\eta$ series. {}More
precisely, the convergence radius of the geometric series in Eq.~(\ref{serIR})
is immediately given by $R^{-1} =1/(24\pi)$, so that, for $u=1$, the original
$u/\eta$ series (before LDE is performed) can only converge if $| {\bar \eta}
| > 1/(24\pi) \sim 0.013$. [One expects similarly that the exact large-$N$
original series in Eq. (\ref{exp20braaten}) should have a convergence radius
very close to this $1/(24\pi)$ value, as we actually checked numerically by
calculating e.g. $R^{-1} =\lim_{i\to\infty} J_{i+1}/J_i$ to sufficiently high
order $i \sim 10^3$]. On the other hand, the exact result should be recovered
for $\eta =0$, thus outside the convergence radius of the original
perturbative expansion. Nevertheless, just as explained with the simple
geometric series in Sec. IV.A above, this is compensated by the fact that the
reorganized LDE series to order $k$, modify perturbative coefficients of order
$n$ with an extra damping factor of $1/\Gamma[1+n/2]$  (see e.g.
Eqs.~(\ref{ldep}), (\ref{ldepinf}) in Appendix B). As a result, this modifies
the convergence radius so that smaller and smaller values of $|\eta| $ can be
reached, which basically explains e.g. the good convergence of the IR series
to the right exact result shown in Table IV. Moreover,
an important remark in view of the more interesting application to the
finite $N=2$ case, is to realize that most results produced by the
families with real positive parts in Tables II-IV give reasonable values
already at very low orders $k\sim 2,3$, being for instance about $10\%$
away from the exact result. We believe this is not at all a coincidence
but simply reflects the faster convergence properties of the
PMS solution within our prescription, as discussed in details with the
simple example of Sec. IV.A, due to the presence of the linear tadpole
term, which implied for the simpler geometric series case
an exact cancellation for $\eta\to 0$ of the first order term.
For the actual IR geometric series the normalization of
the tadpole term relative to the series is however completely different,
but because $|\bar \eta| \to (24\pi)^{-1}$ as the LDE order $p$
increases, one sees immediately that the first term of the expansion of
the tadpole term for $| {\bar \eta}| \to (24\pi)^{-1}$ is $-
N/(96\pi^2)$, so that a similar cancellation still occurs, rendering
again the convergence somewhat faster. 

Based on those general convergence properties considerations, we may
also introduce a very simple criteria to select among the
multiple PMS solutions as illustrated in Tables II-IV: we can consider
the PMS solutions with the smallest $|\bar \eta|$, yet such that
$|\bar\eta|$ is still within the convergence radius
of the relevant series. Note that this criteria has also the advantage
of automatically eliminating the negative $F0$ solution, as it turns out
that the latter always corresponds to the largest (and
in fact, slowly increasing with $k$) $|\bar\eta|$ values,
just like was the case for the simpler example of Sec. IV.A. Indeed, it
is interesting to compare again the exact versus IR approximated $1/N$
series when replacing within their respective LDE expansion to order $k$
this exact value of $\eta = R^{-1}$, instead of the PMS optimized $\bar
\eta$ values. This is illustrated in Table V, where it clearly appears
that the IR approximated series behaves in a somewhat better way than the
exact series. 

Now coming back to the exact large-$N$ series case in trying to better
understand the results in Tables II and III, we can alternatively study
numerically the exact expression (\ref{basic2N}) directly, that is
performing the integration numerically before expanding in $u/\eta$
series. In contrast with the latter expansion having (before LDE is
performed) finite convergence radius as just discussed above,
(\ref{basic2N}) is of course 
defined for arbitrarily small $\eta$. The behavior
of this expression for $u=1$ close to $\eta=0$ is shown in Fig. 2. As
one can see, besides the exact result at $\eta=0$, Eq. (\ref{basic2N})
gets a real minimum $\sim 2.06187$ for very small $\eta \sim
0.00235745$. Now it appears clearly that all the LDE solutions in Tables
II-III are disturbed by the presence of this real minimum very close to
$\eta=0$: namely in the process of reaching smaller and smaller $\eta$
values after the LDE-PMS procedure is applied, the PMS optimization
solutions of the LDE at successive orders can be ``trapped" into the
first minimum reached. In contrast, notice on Fig. 2 that the IR
approximated series has a similar real minimum but located exactly at
$\eta=0$, therefore it is not surprising that the PMS optimized LDE does
converge correctly into this minimum in this case. This explains why the
exact large-$N$ series converges to a value slightly different from
the exact one, and is accordingly a weak point of the procedure. In view
of this it is interesting to briefly compare the latter results with the
alternative prescription such that the LDE-PMS procedure is applied on
the same series, Eq. (\ref{exp20braaten}), but omitting the tadpole term
linear in $\eta$: $-N \eta^*/(4 \pi)=\langle \phi^2
\rangle^{(\delta)}_u- \Delta \langle \phi^2 \rangle^{(\delta)}_c$. This
corresponds to extremizing $\Delta \langle \phi^2 \rangle^{(\delta)}_c$
directly (see Eq. (\ref{kappa})). This is then similar to the recent LDE
convergence studies of Braaten and Radescu \cite{braaten}. Note that in
this case, the additional extremum at very small $\eta$ is
removed. By selecting for simple illustration only the best PMS
solutions ({\it i.e.} the ones whose real parts are the closest to the exact
$1/N$ result $\sim 2.328..$), we obtain for $c_1$ the results shown in Table
VI. As one can see, the results in this procedure do converge to
the right result, however very slowly. On the other hand, one can also
realize that the results at lowest orders $k \sim 3-4$, are far away
(more than a factor of 2) from the exact result. In fact, one should
wait until about $k \sim 50$ to have a reasonably good approximation.
(Moreover, we note that may other solutions that are not shown in Table VI
start to be very unstable at such high orders, some being several orders of
magnitude away from the correct result). So, even if the series 
without tadpole term ultimately converges to the right result
within this LDE prescription, it
appears of not much practical use for the non-trivial $N=2$ case, where
we recall that only the first three perturbative coefficients are known
at the moment. This is to be contrasted with the results of Tables
I-IV, where the fact that the lowest orders are already a good
approximation is, as above explained, a consequence of the
cancellation of the first order ${\cal O}(\eta)$ term, and of the
behavior of the PMS solutions
as the LDE order $k$ is increased. 

We will next discuss a method which fully exploits these scaling
properties of the LDE-PMS solution at large orders, and which accordingly
allows directly to resum the LDE perturbative series and to eventually
further accelerate the LDE convergence, when applied to our BEC problem.

\subsection{Contour Integral Accelerated LDE Resummation Technique (CIRT)}

Having performed the usual LDE interpolation, with $\eta^* = \eta \sqrt{1 -
\delta} $ and $u \to \delta u$, one obtains the physical quantity
$\Phi$ expanded to order $k$. This procedure defines a partial sum that
may be written as (see Eqs. (\ref {exp20}) and (\ref {exp20braaten}) )

\begin{equation}
\Phi^{(k)}(\eta^*, \delta u) \equiv \;\sum^k_{n=0}\: 
c_n (u \delta)^n [f(\eta^*)]^n \;,
\end{equation}
where $f(\eta^*)$ is a function of $\eta^*$ whose form depends on the
dimensionality of the physical quantity $\Phi$. As we have already
emphasized, the use of $\eta^*$ is just an economical way to take into
account the simple $\delta \eta^2$ vertex. Usually, one expands $\eta^*$
so that all the terms of order $\le k$ are present and the direct
application of the PMS optimization or {}Fastest Apparent Convergence to
this quantity, at $\delta =1$, defines the standard LDE. On the other
hand, one could be tempted to improve a perturbative series for which
the highest order term contains $(\delta u)^k$ and then expand $\eta^*$
obtaining a higher order series. This may seem artificial at first
because the contributions of order greater than $k$ would come only from
$\delta \eta^2$ insertions not taking into account, at the same order,
new contributions which arise from the quartic $\delta u$ vertex.
However, by construction \cite{gn2} this procedure helps to accelerate
convergence as it gives in a more direct and simple way the large LDE
order behavior, as studied by the direct ``brute force'' LDE 
method in
Appendix B. One may consider an all order resummation by observing that,
for $\delta\to 1$, the partial LDE series is given formally, from the
simple pole residues, as

\begin{equation}
\Phi^{(k)}(\eta,u,\delta\to 1) =\frac{1}{2\pi\,i}\,\oint
d\delta\,\frac{\delta^{-k-1}}{1-\delta}\,\Phi(\eta,u,\delta)\;, 
\label{cont1}
\end{equation}
where the anticlockwise contour encircles the origin. Now, one performs
a change of variables (see Ref. \cite{gn2} for the original
application of this procedure) for the relevant $\delta\to 1$ limit, 

\begin{equation}
\delta \equiv 1-v/k  \;,
\end{equation}
together with a similarly order-dependent rescaling of the arbitrary
mass parameter, $\eta \to \eta \;k^{1/2}$, where the power $1/2$ is
simply dictated by the form of the scalar mass interaction term, $\eta^2
\phi^2$ in Eq. (\ref{Sdelta}). This rescaling of the mass parameter is
of course consistent with what is obtained by a direct study of the
large $k$ behavior of the standard LDE, see Appendix B. 

{}For $k\to\infty$ this  resummation takes the final form of    
the replacement 

\begin{equation}
\eta^*
 \to \eta v^{1/2}   \;,
\end{equation}
followed by the contour integration 

\begin{equation}
\Phi^{(k \to \infty)}(\delta \to 1) =
\frac{1}{2\pi i}\, \oint \frac{dv}{v}\: \exp(v)
\Phi( \eta^* \to \eta  v^{1/2}) \;,
\end{equation}
where the ``weight" $\exp(v)/v$ originates from

\begin{equation}
d\delta \,(1-\delta)^{-1} \to -dv/v   \;,
\end{equation}
and

\begin{equation}
\lim_{\,k\to\infty}(1-v/k)^{-k-1}=\exp(v)  \;,
\end{equation}
while the original contour was deformed to encircle the branch cut
$Re(v)<0$. Here, one is initially dealing with a power series in
$(\delta u)[\delta u/\eta^*]^{i}$ (cf. e.g. Eqs.~(\ref{phip}), (\ref{phi0}))
and so, the use of

\begin{equation}
\oint dv \exp(v)\: v^a = 2\pi\,i /\Gamma(-a)\,\;,
\end{equation}
shows that the main effect of this resummation is to divide the original
expansion coefficients at order-$\delta^{(i+1)}$ by terms $\Gamma(1+i/2)
\sim (i/2)!$ for large $n$. This damping of the perturbative
coefficients at large order, as implied by this specific resummation, is
fully consistent with what is obtained by a direct
``brute force" resummation of the LDE series for large order $k\to\infty$, 
see e.g. Eqs.~(\ref{ldep}), (\ref{ldepinf}) in Appendix B. But such a damping
is rather generic and was exploited recently in the completely different
context of asymptotically free models \cite{jldamien} where it was shown to
accelerate convergence of the LDE. When applied to the anharmonic oscillator,
it is in fact (asymptotically) equivalent to the more direct LDE resummation
with an order-dependent rescaling of the arbitrary mass, as employed in some
of the Refs.~\cite{ian} to establish rigorous convergence of the LDE for the
oscillator energy levels, that is itself an extension of the order-dependent
mapping (ODM) resummation technique (see, Seznec and Zinn-Justin in Ref.
\cite{early}). In fact, this procedure can essentially suppress the factorial
behavior at large orders of the perturbative coefficients generic in many
theories, and convergence may be obtained even for series that are originally
not Borel summable \cite{ian,jldamien}. The above contour integral
resummation is very convenient since it is algebraically simpler than
the direct LDE summation (compare with Appendix B). In the present case, one
expects fast convergence since the original series has no factorially divergent
coefficients. 

Let us start by treating the standard closed form result
of this contour integral accelerated resummation technique (CIRT)
transforming Eq.(\ref {exp20braaten}) into

\begin{equation}
\langle \phi^2 \rangle_u^{(k)}= - \frac {N \eta}{4\pi \Gamma(1/2)} +
 \frac {N u}{3}
\sum_{i=1}^{k} \frac {J_i}{\Gamma(1+i/2)}   \left ( - \frac { u N
}{6 \eta}
\right )^i   \;,
\label{exp20cirt}
\end{equation}
which clearly displays the coefficients damping. Then, by applying the
PMS, Eq. (\ref{pms1}), one obtains the results displayed in Table VII.
This table shows that only the lowest order real parts of the positive
families produce reasonable results but the values seem to deteriorate
at high orders. Again, one observes that all positive families have
similar behavior, starting from values close to $\sim 2.000$ the values
decrease and then start to increase, as exemplified by $F1$. Table VII
seems to show that the disturbance due to the presence of the extra
minimum of Eq. (\ref{basic2N}): $\sim 2.06187$ for very small $\eta \sim
0.00235745$, reflected in the data of Table II, get amplified by the
CIRT which resums the series.
This effect will be further discussed in Sec. IV.E below.

We shall now examine how the all order LDE summation can further improve
the convergence of the series. To see that, we consider again the IR
behavior (\ref{asymN}), that is sufficient to grasp the essential
features by keeping all results fully analytical. {}Applying thus the
CIRT method to the simpler geometric IR series, Eqs.~(\ref{asymN}) and
(\ref{serIR}), we obtain the result

\begin{equation}
\langle  \phi^2 \rangle_u^{(k)} =- \frac {N \eta}{4\pi \Gamma(1/2)} +
\frac{N u}{96\pi^2}\:\left[\sum^k_{i=0}  \frac{(-x)^i}{\Gamma(1+i/2)}  -1
\right]\;\;=- \frac {N \eta}{4\pi \Gamma(1/2)} +
\frac{N u}{96\pi^2}\:\left[e^{x^2} Erfc(x) -1\right]
\label{Erf}
\end{equation}
with $x\equiv N\,u/(48\pi\eta)$ and $Erfc(x)=1-Erf(x)$ is the standard
error function (see also Appendix B).  
Next, upon using the well known asymptotic
expansion of $Erfc(x)$ for $x\to\infty$ ({\it i.e.} $\eta \to 0$) \cite
{abramowitz} (see Eq.~(\ref{Erfas}) in Appendix B), we can again apply
the PMS optimization procedure at given successive orders $k$. The PMS
optimization generates the results shown in Table VIII. Note how the
large-$N$ result, $c_1=8 \pi /[3 \zeta(3/2)^{4/3}]\simeq 2.328$ is 
{\it exactly} reproduced already at the
lowest non-trivial order. Moreover, as one can see from that table, this
real solution remains valid at any order in perturbation theory. Thus,
the exact large-$N$ result is recovered from this exact CIRT resummation
of the LDE, even if we have only used the IR approximation of the
propagator. The convergence to the exact $1/N$ result in this
alternative LDE implementation is extremely rapid upon using the CIRT.
This also indicates that the convergence to the exact result can be
independent of the details of the non-asymptotic behavior of the
perturbative series coefficients, at least for the large-$N$ quantity
here considered. 

An important feature of the CIRT-PMS results in Table VIII is that the
asymptotic expansion in Eq. (\ref{Erfas}) implies that we are now
dealing with a series in $\eta/u$, instead of the standard perturbative
series in $u/\eta$, which we started from, as in Eqs. (\ref{exp20}),
(\ref{exp20braaten}), or (\ref{serIR}). This is a consequence of the IR
approximation, leading to the simple geometric series in $u/\eta$,
Eq.~(\ref{serIR}), whose all order CIRT form has an exact expression,
Eq.~(\ref{Erf}), that can be re-expanded in a $\eta/u$ series. The
reason why the convergence properties of such a series are much better
than those of the original series should now be clear in view of the
discussion in previous subsections:  the original theory is to be 
recovered in the limit $\eta \to 0$, which is clearly not in the (finite)
convergence domain of the original $u/\eta$ series, while it is
automatically inside the convergence domain of the all LDE order
resummed series Eq.~(\ref{Erf}).
Moreover, another advantage of this reverted series approach is that we
may bypass the need of PMS or other similar criteria: clearly the best
approximation to the exact result will be simply given by the smallest
$\eta$ values, irrespective of whether it is a solution of a PMS (or
similar) criteria.

\subsection{LDE from Pad\'e approximants}

{}For completeness we will also consider in this subsection the results
obtained from a completely different resummation of the relevant BEC
series, based on the Pad\'e approximant (PA) method \cite{pade}. We
define, as is standard, a PA $P[n,m]$ as a rational fraction of two
polynomials of order $n$ and $m$ respectively in the relevant variable
$u/\eta$:

\begin{equation}
P[n,m](u/\eta^*) \equiv \frac{\sum_{i=0}^n a_i (u/\eta^*)^i}
{\sum_{j=0}^m b_j (u/\eta^*)^j} 
\label{pade}
\end{equation}  
where the perturbative coefficients $a_i$ and $b_j$ are obtained order
by order by expanding Eq. (\ref{pade}) up to order $n+m$ and matching
the resulting series with the original expansion. We recall that PA are
generally useful when a series is known only up to the first few orders,
as they can predict sometimes with a very good accuracy the unknown
higher orders, and/or give very good resummation results. As far as the
BEC series is concerned, our further motivation to consider PA is that
it will allow to simply define an alternative (approximated) series in
the {\em inverse} variable $\eta/u$, starting from the exact large $N$
(or finite $N$ as well) series Eq.~(\ref{exp20}). As illustrated in
previous subsection with the IR approximated series that has an obvious
alternative expansion in $\eta/u$, since the exact result is recovered
for $\eta \to 0$, we shall expect much better convergence properties
from such an inverted series, as will be seen below. Note also that the
PA method is largely independent of the previous methods, so that it can
provide a further consistency cross-check of the numerical results. 
Another advantage is that the PA technique is
immediately applicable as well to the finite $N$ case, to be discussed
below in Sec. V. A
drawback of the PA method, however, is that 
they are not uniquely defined, since for a given perturbative expansion 
of order $k$ one may consider a priory all
possible PA with $n+m=k$. In order to thus limit somehow the number of
possible PA without introducing much bias in our analysis, we will only
consider resulting PA that can be expanded in powers of $\eta/u$, for
the reasons discussed above, which imposes that $n <m$. To illustrate in
the most simple case the power of PA, let us first consider again the IR
approximated geometric series Eq. (\ref{serIR}), but assuming that only
the first order term of the $u/\eta$ series is known: $\langle \phi^2
\rangle_{IR} \sim -(N\eta^*)/(4\pi) + (\delta u N)/[(96\pi^2)(8\pi)] (-
\delta u \, N)/(6\eta^*)$. We could then define an approximation of this
series as follows

\begin{equation}
\langle  \phi^2 \rangle_{IR} \sim -\frac {N\eta^*}{4\pi}
+ \frac {\delta u  N}{96\pi^2} [P[0,1](u/\eta^*) -1]
\label{pade1}
\end{equation}  
where the PA of order $n+m = 1$ is

\begin{equation}
P[0,1](u/\eta) \equiv
[b_0 + b_1 \delta\,u N/(6\eta)]^{-1}\;,
\label{pad1}
\end{equation}  
and a simple matching of the expansion of (\ref{pade1}) gives $b_0=1$
and $b_1= 1/(8\pi)$, such that the exact geometric series is in fact
recovered, and can be of course expanded in the alternative form of a
$\eta/u$ series. Though this example maybe too simple,
 we can expect that the more general PA that are constructed
below to approximate the more complicated exact series, will 
have similarly
good resummation properties. Typically at order 3 we have to consider
$P[1,2]$ and $P[0,3]$.  The PA results are shown in Table IX for the 
large $N$ case. The
order $k$ designates in this case the order of the (re)-expansion in LDE
power series of the PA, followed again by a standard PMS optimization.
Higher orders $k >5$ are not shown but exhibit a very stable behavior
with solutions almost similar to the lowest orders shown. As one can
see, the exact result is often reproduced as a PMS solution, which is
not so remarkable in the present $1/N$ case, as it simply means that
$\eta=0$ (for which value the PA is well-defined by construction) is a
solution when applying the PMS to the PA. Clearly the best PMS
solutions, if $\eta \neq 0$, are obtained from the smallest $|\eta| \neq
0$ values. Though it appears clearly that the PA $P[0,3]$ give much better
results than the $P[1,2]$, note that
the latter give solutions $\sim 2.05$
and $\sim 2.6$ that are located not far below and above the exact
value. The better $P[0,3]$ results maybe 
eventually explained by noting that
the $P[0,3]$ should have 
much better resummation properties for
$\eta\to 0$, having no numerator term in $u/\eta$. 

Another 
advantage of the PA is that one may skip the PMS criteria and simply
take the limit $\eta\to 0 $ directly, in which case the exact $1/N$
result is reproduced. Of course as concerns the large-$N$ case those
results are only a consistency cross-check, 
since we started from the exact series anyway
and thus only managed to define a PA such that its $\eta\to 0$ limit is
well-defined, in contrast with the naive $u/\eta$ series. But the very
same PA procedure can be applied to the finite $N$ case to be discussed
in the next Section.

\subsection{Brief summary and discussion of the convergence properties}

One may now summarize the main features of this detailed large-$N$
investigation so that the finite $N$ application, in the next Section,
can be carried out straight away. {}Regarding the family selection one
notes, especially in Tables II and IV, that except for $F0$ all families
produce results that converge to approximate similar values at high
orders, as in the anharmonic oscillator an geometric series cases.
Therefore, choosing consistently $F1$ in all the BEC applications
appears to be an appropriate choice since this family is also the only
one which allows predictions at any order ({\it e.g.} in a computation
involving only low orders in $\delta$). Again, this is consistent with
the observations drawn from the geometric series analysis.

To understand completely the reason for convergence it is particularly
convenient to examine the formal expressions of the large LDE order
behavior, i.e. for $k\to\infty$, as derived in details in Appendix B.
What happens, as already explained in Sec. IV.B, is
that when the LDE order is
increased, the numerical 
PMS optimal solutions $|{\bar \eta}|$ tend to be smaller and smaller and thus
to reach the boarder of the convergence radius of the original perturbative
series. But at the same time, from the study of the asymptotic behavior of the
series at large LDE orders we see that the main effect of the rearranged
LDE series is to provide an extra damping factor $1/\Gamma[1+n/2]$ 
 as well as a scaling factor $k^{n/2}$ in its
order $n$ coefficient.   So, 
after redefining conveniently $\eta \to \tilde \eta k^{1/2}$,
the {\em new} optimized $\tilde \eta$ values can tend to zero, for
which the new LDE-resummed series is now more and more convergent (see {\it
e.g.} Eq.~(\ref{Erf})) and leads to the
correct $\eta\sim 0$ result. 
However, in the process of reaching smaller and smaller
values, it may happen that another non-trivial extremum, if present, is
first met, in which case the convergence is disturbed or slowed
down by the presence of
this other extremum and the $\eta\to 0$ may no longer be reached. This
is precisely what happens in the case of the exact $1/N$ series, as
illustrated above in Tables II and III, 
with this problem worsening
as the series is resummed, as shown by the results in Table VII, for the CIRT
resummation applied to the exact large-$N$ perturbative series. 
In contrast, by using  the
IR series approximation the disturbing extremum that now
approaches $\eta  \neq 0$
is removed and this problem is not present there.
{}Finally, we mention that the recent LDE convergence studies of Braaten
and Radescu \cite{braaten} follow the same lines of the standard LDE
calculations performed by us in Sections III.B and IV.B. The major
difference being that those authors prefer to extremize $\Delta \langle
\phi^2 \rangle^{(\delta)}_c$ directly (see Eq. (\ref{kappa})). In
practice, the difference in between those physical quantities amounts to
the tadpole term $-N \eta^*/(4 \pi)=\langle \phi^2 \rangle^{(\delta)}_u-
\Delta \langle \phi^2 \rangle^{(\delta)}_c$, that appears in Eqs. (\ref
{exp20}) and (\ref {exp20braaten}). In the case of the standard LDE-PMS
application, like the one shown in Table II, their numerical results are
similar to ours at very large orders, exhibiting
ultimately convergence, but do not give a good
approximation at the lowest orders. The reason for this faster
convergence when the tadpole term is present is the cancellation of the
leading terms for $\eta\to 0$ as explained in detail above. Our simple
geometric series investigation has also shown the crucial role,
regarding convergence, played by this type of linear term. Moreover,
these type of loop terms are also at the origin of good convergence
properties observed in many other applications \cite {phil, sunil,
gastao}. 

\section{ THE FINITE $N$ CASE WITH RESUMMATION OF THE LDE}

Let us now turn our attention to the finite $N$ results and, especially,
to the improvement of the LDE within this limit by using the CIRT
resummation method discussed in the previous Section. The insight gained
in the detailed large-$N$ study will prove to be very useful to
understand the structure of optimized results and to select the
appropriate solutions. {}For arbitrary $N$, the quantities $\langle
\phi^2 \rangle_u^{(k)}$ as well as $\delta r_c^{(k)}$ have been
evaluated in details, up to order-$\delta^4$, in Ref. \cite {pra}. 
The contributing diagrams evaluated in that reference for the $\delta$
perturbative expansion of $\langle
\phi^2 \rangle_u^{(k)}$ to order-$\delta^4$
are shown in 
{}Fig. 3. In Ref. \cite{pra} all the integrals appearing
in those diagrams were
obtained with the type of perturbative calculation discussed in Subsec.
III.A with the multidimensional {}Feynman integrals calculated with
VEGAS. Those terms were 
also later been obtained  by Braaten and Radescu
\cite{braaten} in a different way, by reducing the multidimensional
integrals to one-dimensional ones in some of the terms of fourth order in
$\delta$ and, more recently, Kastening in Ref. \cite{boris}
has  also revised those numerical results, obtaining 
more precise numerical results for the integrals. {}From the results
originally obtained in Ref. \cite{pra} and using the corresponding
updated, higher precision coefficients evaluated in \cite{boris}, 
the terms contributing to  $\langle
\phi^2 \rangle_u^{(k)}$ to order-$\delta^4$, shown in {}Fig. 3,
are given by

\begin{eqnarray}
\langle \phi^2 \rangle_u^{(4)} &=&
- \frac{N\eta^*}{4\pi}
-\delta^2 \frac {u^2}{\eta^*} \frac {N(N+2)}{18(4\pi)^3} [0.14384] 
\nonumber \\
&+& \delta^3 \frac {u^3}{(\eta^*)^2} \frac{N}{(4\pi)^6}
\frac{\left(16+10 N + N^2 \right)}{108} [8.06940]
 + \delta^4
\frac {u^4}{\eta^3} \frac {N(N+2)^2}{ (18)^2 (4\pi)^7}[0.11507]
\nonumber \\
&-&\delta^4 \frac {u^4}{\eta^3} \frac{N}{(4 \pi)^7}
\frac{\left(40 + 32 N + 8 N^2 + N^3 \right)}{648} [3.12811] - \delta^4
\frac {u^4}{\eta^3} \frac{N}{(4 \pi)^7}
\frac {\left( 44 + 32 N + 5 N^2 \right)}{324} [1.71859]
\nonumber \\
&+& \delta^4
\frac {u^4}{\eta^3} \frac {N(N+2)^2}{108 (4\pi)^7}[0.20821]
-\delta^4 \frac {u^4}{\eta^3} \frac{N}{(4 \pi)^7}
\frac{\left( 44 + 32 N + 5 N^2 \right)}{324}
[2.66746] +  {\cal O}(\delta^5) \;.
\label{first3}
\end{eqnarray}
where, as before, $\eta^*=\eta\sqrt{1-\delta}$ must be expanded
accordingly to take into account all contributions shown in {}Fig. 3. 
{}The symmetry factors appearing in Eq. (\ref{first3}) can be found
e.g. in \cite{kleinertsym}. The
symmetry coefficients clearly show that this perturbative expansion is
valid for any $N$ which means that, up to order-$\delta^4$, the
large-$N$ results Eqs. (\ref {exp20}) and (\ref {exp20braaten}) may be
recovered, within numerical error bars of about $ 2\%$. This evaluation
is easily done by considering $u$ to be of order $1/N$ so that only
$(uN)^k$ terms are retained together with the first, $u$-independent
term. {}Fig. 3 illustrates well how the LDE mixes, at a given order,
diagrams which normally appear at different orders in the $1/N$
expansion.

By setting $N=2$ one then gets the more compact form

\begin{equation}
\langle \phi^2 \rangle_u^{(4)}= - \frac { \eta^*}{2\pi} + \delta u
\sum_{i=1}^{3} K_i \left ( - \frac {\delta u}{ \eta^*} \right )^i  +
{\cal O}(\delta^5) \;\;,
\label{expNF}
\end{equation}
where the coefficients are $K_1=3.22158 \times 10^{-5}$, 
$K_2= 1.51792 \times  10^{-6}$, $K_3= 9.66514 \times 10^{-8}$.
It is worth remarking 
that the coefficients $K_1$ and $K_2$ obtained from the numerical
results of Ref. \cite{pra} agree with these results, also obtained
later by the authors of Refs. \cite{braaten,boris}. At the same time the 
$K_3$ coefficient used here, which was obtained in Refs. \cite{braaten,boris},  
differs by about $ 10 \%$ from the one
that would come from the results of Ref. \cite{pra}. 
We in principle can 
trace this difference by the fact that
five non-trivial graphs with five loops enter the
evaluation of $K_3$ (see {}Fig. 3), which require some careful
calculation\footnote{We thank B. Kastening for pointing out to
us the correct values for the five-loop diagrams and for 
discussions concerning the difficulties in evaluating them.}.

Then, by applying the standard PMS,
Eq. (\ref{pms}), to Eq. (\ref{expNF}) one obtains three families, 
shown in Table X, in
agreement with Ref. \cite{pra}. Turning to the CIRT resummation of Eq.
(\ref {expNF}) one proceeds as in the large-$N$ case (see Sec. IV.C).
By applying the CIRT improved in the context of the PMS optimization to
Eq. (\ref {expNF}) one obtains three other families, also shown in Table
X. In fact, the previous large-$N$ analysis strongly suggest that also
here the first family with positive real parts should be the relevant
family concerning our $N=2$ predictions. Indeed, if we assume that the
large order behavior of the actual $N=2$ series coefficients should not be
drastically different from the analogous large-$N$ series, all the fast
convergence and scaling properties that were
discussed in Sec. IV should be approximately valid also for $N=2$. Then,
the first negative family in Table X is easily eliminated by the same
criteria as in the large-$N$ case, because it again always corresponds
to the largest $|\bar\eta|$ and does not exhibit any trend towards smaller $|\bar\eta|$ 
values as the LDE order $k$ is increased. Similarly, we also notice that the
$F_1$ family in Table X has ${\rm Re}(\bar\eta)$ substantially smaller than
$F_2$, while we expect the exact result to be for $\eta\to 0$. Moreover, due
again to the presence of the tadpole term in our procedure, from the
analysis in Sec. IV we can expect that our results, though intrinsically
limited for $N=2$ to the first four LDE orders, should be nevertheless
already a reasonably good approximation.

A further crosscheck of the consistency of our results without any
knowledge of the exact higher order coefficients for the case $N=2$, is
the stability of the result when replacing these unknown perturbative
coefficients by a well-defined approximation. This is the result
exhibited in Table XI, where the unknown order $K_i$ with $i \ge 4$ were
replaced by the corresponding coefficients of the IR approximated
series, Eq.~(\ref{serIR}). As one can see, the stability of those
results is quite remarkable, and one can even observe the slow
convergence of the standard LDE-PMS to the CIRT result. 

The physically meaningful real part of our order-$\delta^4$ CIRT
improved result $c_1\simeq1.19$ can then be compared with the recent Monte
Carlo estimates $c_1=1.32 \pm 0.02$ and $c_1=1.29 \pm 0.05$. Note that
the standard order-$\delta^4$ PMS result, $c_1 \simeq 1.53$ is also
a satisfactory estimate. We note that the CIRT and ordinary PMS results
just bound the Monte Carlo estimates from below and above, respectively.

{}Finally, in Table XII we show the results for $c_1$ obtained by
Pad\'{e} approximants (PA), as discussed in the last Section. As
expected, at first non trivial LDE order (order 3), only the $1/N$
solution is found from the PA, because only $\eta=0$ is a PMS
optimization solution. As one can see, the non-trivial results at higher
orders (5--10) of the LDE expansion in Table XII are nicely consistent
with what is independently obtained from the standard LDE and CIRT
results shown in Table X. Note however that the PA $P[0,3]$ are in
better consistency than the $P[1,2]$, which in fact only give result
very similar to the ones in the large-$N$ case shown in Table IX. \\

The
contributions to $\delta r_c^{(4)}= - \Sigma_{\rm ren}^{(4)}(0)$, which
enter in the derivation of the constant $c_2^{\prime \prime}$, Eq. (\ref{c2}),
have
also been explicitly evaluated in Ref. \cite {pra} and again we refer the
interested reader to that reference for the details and we show below
only the final result for the renormalized, scale ($M$) dependent $\delta r_c^{(4)}$
obtained in that reference:

\begin{eqnarray}
-\delta r_c^{(4)} &=& \Sigma^{(4)}_{\rm ren}(0)=   -
\delta u \frac{\eta^*}{8 \pi}\left ( \frac{N+2}{3} \right )
-\delta^2 \frac{u^2}{(4\pi)^2} \frac{(N+2)}{18} \left [
\ln \left ( \frac{M}{\eta^*} \right ) - 0.59775  \right ]  \nonumber \\
&-& \delta^3 \frac {u^3}{\eta^*} \frac {(N+2)^2}{108( 4
\pi)^3}[0.143848]    +\delta^3 \frac {u^3}{\eta^*}
\frac{\left( 16 + 10 N +  N^2 \right)}{(4\pi)^5 108}  [81.076]\nonumber \\
&+&\delta^4 \frac {u^4}{\eta^2} \frac {(N+2)}{6 (4\pi)^6}
\frac{\left( 16 + 10 N +  N^2 \right)}{108}    [8.09927] \nonumber \\
&-&
\delta^4 \frac {u^4}{\eta^2}
\frac {\left( 40 + 32 N +8 N^2 + N^3 \right) }{(4\pi)^6 648}  [20.43048]
- \delta^4 \frac {u^4}{\eta^2}
\frac {\left( 44 + 32 N + 5 N^2 \right) }{(4\pi)^6 324}  [12.04114]
\nonumber \\
&-&
\delta^4 \frac {u^4}{\eta^2}
\frac { \left( 44 + 32 N + 5 N^2 \right) }{(4\pi)^6 324}
[17.00434]
+ \delta^4 \frac {u^4}{\eta^2} \frac {(N+2)^2}{(18)^2(4\pi)^6}
[2.8726]+  {\cal O}(\delta^5) \;,
\label {rc}
\end{eqnarray}
which, for $N=2$, becomes

\begin{equation}
\delta r_c^{(4)}= - \Sigma_{\rm ren}^{(4)}(0)=  \delta \frac {u \eta^*}{6\pi} 
+ \delta^2 u^2 A_2 \left [ \ln
\left ( \frac{M}{\eta^*} \right ) - 0.59775 \right ] 
-\delta^3 \frac {u^3}{\eta^*} A_3 +
\delta^4 \frac {u^4}{(\eta^*)^2} A_4  +  {\cal O} (\delta^5) \;\;,
\label{rcNF}
\end{equation}
where $A_2=1.40724 \times 10^{-3}$, $A_3=8.50859 \times 10^{-5}$ and
$A_4=3.52299 \times 10^{-6}$. The application of the PMS optimization to
Eq. (\ref{rcNF}) reproduces the same results as obtained in Ref.
\cite{pra} which are $\sim 101.4$, $\sim 98.2$ and $\sim 82.9$ from
second to fourth order respectively.

At the same time, treating $\delta r_c^{(4)}$ with the CIRT
(from Sec. IV.C) one obtains
the result\footnote{Note that the scale $M=u/3$ was originally chosen in
the Monte Carlo applications \cite {second}.} ${\rm Re} \;[
r_c^{(4)}(M=u/3)]=0.0010034 u^2$ which, together with the CIRT improved
$\langle \phi^2 \rangle_u^{(4)}$ result, and Eq. (\ref{c2}), leads to
(with errors estimated from the integrations performed in \cite{pra} with Vegas)
$c_2^{\prime \prime}= 84.9 \pm 0.8$ whereas the Monte Carlo result is
$c_2^{\prime \prime}= 75.7 \pm 0.4$ \cite {second}. To our knowledge,
these are the only analytical predictions for this coefficient to the
present date. For consistency, note that the optimization of $\delta
r_c^{(4)}$, including the selection of solutions, has also been
performed according to what was done for $\langle \phi^2
\rangle_u^{(\delta)}$.

Note that we have not attempted to examine the infrared behavior of the
finite $N$ case since its series is much more complicate than the
large-$N$ one. However, our previous large-$N$ investigation shows that
the LDE works well, even for the standard series, already when considering
only the lowest order terms. 
Here, only these lowest order terms were computed so that
our results, up to order-$\delta^4$, can be considered good estimates
even if one knows that the whole procedure may get spoiled at very high
orders.

\section{CONCLUSIONS}

We have investigated how the LDE followed by a standard PMS
optimization performs in the non-trivial case of phase transitions of
interacting homogeneous dilute Bose gases described by an effective three
dimensional $\phi^4$ field theory. This nonperturbative method has been
recently employed in Refs. \cite {prb,pra} and in Ref. \cite{braaten} to
determine the critical temperature for such a system, giving good numerical
results. One advantage is that the formal calculations are performed exactly
as in the perturbative case. This means that, at each order, one deals with a
very reduced number of contributions which are not selected according to
their topology (like the number and type of loops). Therefore, the
method is valid for any finite value of $N$. To handle ultraviolet
divergences, the renormalization program is implemented as in the usual
perturbative way. Also, an arbitrary mass parameter consistently
introduced by the method avoids any potential infrared problems. 

The convergence properties, including rigorous proofs, of those
nonperturbative methods have been studied in quantum mechanics
\cite{ian,kleinert2,guida,phil} and more recently in quantum field
theory \cite{jldamien}. However, despite the many successes obtained
with the LDE in different applications, the convergence study in the BEC
case poses new challenges. One of the reasons is that it is difficult to
establish simple analytical links between the LDE and other
nonperturbative methods at the one-loop level, since those terms do not
contribute at the transition point. This could rise the suspicion that
the results obtained in Refs. \cite{prb,pra}, for the realistic $N=2$
case, are just a numerical coincidence. The exact value for the linear
coefficient $c_1$ which appears in the critical temperature of
interacting homogeneous dilute Bose gases is still unknown \cite
{markus} although much progress has been recently made concerning its
determination \cite{arnold1}. Here, our aim was to prove the reliability
of the recent LDE results for $c_1$ and $c_2^{\prime \prime}$ \cite{pra}
through a detailed analysis of the LDE convergence properties.

We have started our convergence study by considering the effective BEC
model in the large-$N$ where the results obtained by Baym, Blaizot and
Zinn-Justin ($c_1 \simeq 2.328$) \cite {baymN} can be considered ``exact". We
have performed the usual perturbative LDE evaluation of $\langle \phi^2
\rangle^{(\delta)}_u$ in different ways which allow for numerical
accuracy checks. By considering the asymptotic infrared behavior of the
propagator we have obtained a simpler series with exact coefficients
which allows for a fully analytical analysis. Before tackling the
optimization of the BEC series we have investigated how the procedure
works considering a simple geometric series. The insight gained during
this exercise proved to be very important in understanding the
family structure of optimal solutions with regard to convergence
properties. Then, the standard series and its infrared limit have been
optimized with the PMS criterion, Eq. ({\ref{pms}), leading to reasonable
results in the standard case (see Table II). At the same time, the
numerical results produced by the same optimization applied to the
infrared series display better convergence properties, as shown in Table
IV. This also shows that the optimization procedure is rather sensitive to the
actual form of the perturbative expansion series
coefficients, in particular at low perturbative orders. Indeed, we emphasize
again that including the linear tadpole term in the LDE-PMS procedure
is crucial for a faster convergence and to obtain an already very good
approximation at low perturbative orders, though both procedures tend to
similar results at very high orders. 

We have then presented an efficient all order resummation technique,
similar to the one
used to prove the LDE convergence within quantum field theories at zero
temperature \cite {jldamien}. This LDE resummation method takes advantage
of contour integration techniques which allows to resum more directly the
series and thus to accelerate convergence. Applying this 
contour integral resummation technique (CIRT)  to the exact large-$N$
perturbative LDE series seems however at first to amplify the numerical
instabilities generated through 
numerical optimization. This problem becomes more severe as one moves to
higher perturbative orders as shown in Table VII. On the other hand,
applying the CIRT to the infrared series {\it exactly} reproduces the
large-$N$ value $c_1=2.328$ already at the first non-trivial order. This
solution has no complex parts and remains valid as one goes to higher
orders while all other families of solutions display good convergence
properties having real parts that are numerically very close to the
exact value (see Table VIII). In summary, this extensive analysis has
shown that, for this type of series, the optimization procedure 
and convergence rate may be
influenced by the actual values of the first few perturbative order
coefficients. We have shown that, for this effective BEC model, the simple infrared
series retains all the nonperturbative information and that the LDE, augmented
with the CIRT, performs rather well. Finally, we have seen that all families,
display a very similar structure and will predict approximately the
same values at high perturbative orders. We also have shown how the family
which produces negative $c_1$ values is easily eliminated because it does not
correspond to the expected trend towards small $|\eta|$ values of the PMS
solution. The very same criteria allows to single out
the values generated by the first family of positive solutions, both for the
large-$N$ or finite $N$ case. {}Finally, the results obtained
by a different resummation based on Pad\'e approximants, aimed as an
alternative
to define the relevant $\eta\to 0$ limit from the perturbative series
in $u/\eta$ prior to LDE, appear quite consistent with the latter ones.

More formally, our present study investigated in some  detail 
the large order behavior of the LDE (e.g. in Sec. IV.C with the CIRT
method
or alternatively in Appendix B), from which we can also point out interesting
analogies as well as differences with the LDE convergence properties  in
quantum mechanics for the anharmonic
oscillator\cite{ian,kleinert2,guida,phil}.  The latter is described by a
(one-dimensional) scalar theory with a $\phi^4$ interaction term, and as is
well-known its energy levels have perturbative expansion coefficients that are
factorially growing at large orders\cite{ao}. Nevertheless, as already
mentioned in Sec. II, the LDE can converge essentially because the PMS
optimized solutions behave like a rescaling of the mass parameter with
perturbative order which can compensate the factorial behavior at large
orders. In contrast, the relevant BEC perturbative series here considered in
Sec. III have a finite convergence radius, such that no explicit
rescaling of the mass parameter should be necessary in principle for
convergence. Nevertheless, what the LDE followed by PMS optimization is
performing is to enlarge the original series convergence radius, 
thus qualitatively similarly in this respect with the
oscillator. These properties of the LDE are best exploited by the CIRT more
direct resummation method.   

The final part of the work was devoted to the realistic finite $N$
case for which only the standard series with coefficients numerically
obtained is available. In practice, here only the first low orders
contributions could be evaluated and the comparisons performed in the
large-$N$ show that VEGAS produces, in this case, accurate coefficients
which should not completely spoil the optimization procedure. For
consistency with the large-$N$ case we have considered only the real
parts of the first family of positive solutions as the relevant ones.
Applying the CIRT to this case has improved the recent order-$\delta^4$
results of Ref. \cite {pra}, generating $c_1 \simeq 1.19$ and
$c_2^{\prime \prime} \simeq 84.9$, which are about $9 \%$ smaller  
and $11 \%$ higher, respectively, than the recent lattice Monte Carlo
estimates \cite {arnold1,russos}. In any case one cannot expect to make
a definitive analytical prediction, for those coefficients, from a
calculation involving only a handful of contributions. Nevertheless, the
agreement between our improved analytical LDE results and the recent
numerical Monte Carlo results is quite impressive. 
Moreover, 
the consistency of our $N=2$ results as obtained from a different 
resummation
method (Pad\'e approximants) used prior to the LDE procedure is also
noticeable. Our results seem to
support the fact that, analytically, the leading contributions to
$\Delta T_c$ for the BEC case studied here can be obtained by resumming
typical leading and next to leading $1/N$ type of graphs as in Refs.
\cite {arnold} and \cite {pra}. The better LDE numerical values may be
due to the mixing of such contributions since in reality, $N=2$.

In summary, our detailed convergence study together with the improved
optimization procedure results show the potential of the LDE to tackle
nonperturbative calculations in field theory at critical points. We have
explicitly shown how meaningful non perturbative results for the BEC
problem can be obtained in a consistent fashion which also works for a
general case such as the simple geometric series analyzed in the paper.

\acknowledgments

M.B.P. and R.O.R. were partially supported by Conselho Nacional de
Desenvolvimento Cient\'{\i}fico e Tecnol\'ogico (CNPq-Brazil). M.B.P.
also acknowledges the LPMT in Montpellier for a CNRS {\it Poste Rose}
grant. The authors thank Gilbert Moultaka for discussions related to
Sec. IV.

\appendix

\section{REVIEW OF THE ORIGINAL LARGE-$N$ CALCULATION.}

Let us briefly recall how the exact large-$N$ derivation of $c_1$ was
performed in the original calculation, Ref. \cite{baymN}, in order to
exhibit the differences with the LDE (exact or approximated) evaluation
as performed in Sec. III. Considering the original theory described by
Eq.(\ref {action2}) at the critical point, the Hugenholtz-Pines (HP)
theorem imposes \cite{baymprl,arnold1} $r_c= -\Sigma(0)$, where
$\Sigma(0)$ represents the field self-energy with zero external
momentum. Then, one has a massless propagator and the appearance of IR
divergences has to be carefully dealt with. In Ref. \cite{baymN}, after
applying the HP theorem, the relevant expression for $\Delta \langle
\phi^2 \rangle$ reads

\begin{equation}
\Delta \langle  \phi^2 \rangle =
\langle  \phi^2 \rangle_{u} -\langle  \phi^2 \rangle_{0}
= N \int \frac{d^3\,p}{(2\pi)^3} \;
\left[ \frac{1}{p^2 +\Sigma(p)-\Sigma(0)} \;-\frac{1}{p^2}  \;\right]\;,
\label{basm0}
\end{equation}
where $1/{p}^2$ represents the term with no interaction,
$\langle  \phi^2 \rangle_{u \to 0}$, to be subtracted according to the
discussion in Sec. II (compare e.g. with Eq. (\ref{kappa})), and

\begin{equation}
\Sigma(p) \equiv \frac{2}{N} \int \frac{d^3\,k}{(2\pi)^3}\;
\;F(k)\;\frac{1}{(k+p)^2}  \;\;,
\label{sigma}
\end{equation}
with the ``dressed" (resummed) scalar propagator (see Fig. 1)

\begin{equation}
F(k) = \left [\frac{6}{N\,u} +B(k) \right]^{-1}   \;\;,
\end{equation}
where

\begin{equation}
B(k) = \int \frac{d^3\,q}{(2\pi)^3} \;\frac{1}{q^2\;({ k} +{ q})^2}   \;
=\frac{1}{8\:k}\;,
\end{equation}
represents the basic one-loop (massless) bubble integral.
{}At the relevant (next-to-leading) $1/N$ order one obtains the basic
expression to be evaluated as

\begin{equation}
\Delta \langle  \phi^2 \rangle = -2\;\int \frac{d^3\,p}{(2\pi)^3} \;
\frac{d^3\,k}{(2\pi)^3} \;
\frac{1}{p^4}\; \left (\frac{6}{N\,u} +B(k) \right )^{-1}
\left [\frac{1}{(k+p)^2} -\frac{1}{k^2}\;\right ]   \;\;,
\label{BBZ}
\end{equation}
where the $1/p^2$ subtraction term cancels out with the first order term
belonging to the $1/N$ expansion of the term $[p^2
+\Sigma(p)-\Sigma(0)]^{-1}$, in Eq. (\ref{basm0}). Remark that, after
applying the HP theorem, all loop integrals in (\ref{BBZ}) involve
massless scalar propagators. Accordingly, both integrals in
Eq.~(\ref{BBZ}) are IR divergent at intermediate steps, although the
final physical result $\Delta \langle \phi^2 \rangle $ should be an IR
(and UV) convergent quantity. Also, the integral over $k$ is not
(absolutely) UV convergent: it has superficially a logarithmic UV
divergence. Thus, as emphasized in Refs. \cite{baymN,arnold}, one should
be very careful to correctly regularize these integrals before doing
standard manipulations, like typically exchanging the order of the two
integrations in Eq.~(\ref{BBZ}). {}The authors of Ref. \cite{baymN}
chose to work in dimensional regularization, which takes care of both,
UV and IR divergences. We thus now summarize
the main steps of this calculation. First, one integrates over $p$, so
that the second term of the bracket in Eq. (\ref{BBZ}) vanishes, since $\int
d^d p /p^4=0$ in dimensional regularization. The integration of the
non-vanishing first term in the bracket of Eq. (\ref{BBZ}) gives a result
whose behavior for $d \to 3$ is essentially given by $\sim 1/[\Gamma(d-3)]\;$,
that would, naively, give zero as a result. However, it combines with a simple
pole in $d-3$ given by the next $k$ integral (see below). More precisely,
integration of Eq. (\ref{BBZ}) gives (omitting factors that are regular
for $d\to 3$):

\begin{equation}
\int \frac{d^d\,p}{(2\pi)^d} \;
\frac{1}{p^4\;({k} + {p} )^2} = \frac{1}{(4\pi)^{d/2}}\;
\frac{\Gamma(d/2-1)}{\Gamma(d-3)}\;
\frac{\pi}{\sin(\pi d/2)}\; k^{d-6}   \;\;,
\label{lint}
\end{equation}
where the space-time dimension $d $ is kept arbitrary, for the moment.
{}One has next to deal with an integral over $k$ of the generic form
(omitting again nonessential constant overall factors)

\begin{equation}
I_{\phi^2} \sim \int \frac{d^d\,k}{(2\pi)^d} \;
\frac{k^{d-6}}{\frac{6}{N\,u} +B(d)\,k^{d-4}} \;\;,
\label{pint}
\end{equation}
where
 $B(d=3) = 1/8$.
Next, to evaluate Eq. (\ref{pint}), we make the following change of variable:
$P^2 = k^{-\epsilon}$ where $\epsilon \equiv 4-d$, which gives

\begin{equation}
d^d k \to \frac{-2}{\epsilon} p^{-d\,(1+2/\epsilon)}\;d^d P \;\;.
\end{equation}
Then using the basic dimensional regularization formula:

\begin{equation}
\displaystyle
\int \frac{d^d P}{(2\pi)^d}\; \frac{(P^2)^b}{(P^2+R^2)^a}
= \frac{1}{(2\pi)^{d/2}}
\frac{\Gamma(d/2+b)}{\Gamma(d/2)\:\Gamma(a)}\;
\Gamma(a-b-d/2)\;(R^2)^{(d/2+b-a)}  \;\;,
\end{equation}
we
obtain after some algebra 
 
\begin{equation}
I_{\phi^2} \sim \frac{2\pi}{(4\pi)^{d/2}\;
(4-d)\;\Gamma(d/2) \sin(2\pi (3-d)/(4-d))}\;
B^{2(d-3)/(4-d)}\;\left (\frac{6}{N\,u}\right )^{\frac{2-d}{4-d}} \;\;,
\label{AintBBZ}
\end{equation}
where, for $d\to 3$, $\sin^{-1}[2\pi (3-d)/(4-d)]$ combines with
$\Gamma^{-1}(d-3)$ to give a finite result\footnote{Note a misprint in
Eq.~(27) of ref.\cite{baymN}, where the relevant term $\sin [2\pi
(3-d)/(4-d)]$ reads $\sin [\pi (d-2)/(4-d)]$.} proportional to $u$.
Putting back all overall factors, one finds

\begin{equation}
\Delta \langle  \phi^2 \rangle = -\frac{ N u}{96\pi^2}  \;\;,
\label{exactN}
\end{equation}
which is simply related to $c_1$ via Eq.~(\ref{c1}),
in agreement with Ref. \cite{baymN}.

\section{Large order behavior of standard LDE and PMS}

In this appendix we briefly analyze the large order behavior of the
standard LDE, in order to exhibit some generic properties of the LDE-PMS
optimization solutions, as well as the link with the more direct CIRT
method considered in Sec. IV.B.
  
Considering either Eq. (\ref{exp20braaten}) or (\ref{serIR}) with
$\eta^* \equiv \eta (1-\delta)^{1/2}$, the result of its expansion to
order $k$ in $\delta$ followed by $\delta \to 1$ can be written formally
as (the coefficients $K_n$ here refer indistinguishably to either the
$J_n$ of Eq. (\ref{exp20braaten}) or the $G_i$ of Eq. (\ref{serIR}))

\begin{equation}
\Delta \langle  \phi^2 \rangle^{(k)}(\eta,u) =
-\frac{N \eta}{4\pi} \:\sum^k_{n=0} (-1)^n \frac{\Gamma[3/2]}
{n!\:\Gamma[3/2-n]}
+\frac{u N}{3} \sum^k_{n=1} K_n \left(-\frac{u N}{6 \eta}\right)^n\:
\sum^{k-n-1}_{q=0} (-1)^q \frac{\Gamma[1-n/2]}{q!\:\Gamma[1-n/2-q]}
\end{equation}
where the last sum originates from the expansion of $(1-\delta)^{-n/2}$
and the upper limit of this sum takes into account that at 
order $k$ of the $\delta$ expansion, there is a term coming from 
$u  (u/\eta)^n \to  \delta^{n+1} \:u  (u/\eta)^n $
and a term from $(1-\delta)^{-n/2}$. The sums can be performed
analytically to give

\begin{equation}
\Delta \langle  \phi^2 \rangle^{(k)}(\eta,u) =
-\frac{N \eta}{4\pi} \sqrt{\pi} \frac{(-1)^k}{\Gamma[1/2-k]\,\Gamma[1+k]}
+\frac{u N}{3} \sum^k_{n=1} K_n \left(-\frac{u N}{6 \eta}\right)^n\:
 \frac{\Gamma[-n/2+k]}{\Gamma[-n+k]\,\Gamma[1+n/2]}
\label{ldep}
\end{equation}
The expression (\ref{ldep}) as it stands is particularly convenient to
be optimized with respect to $\eta$ at arbitrary order $k$, leading to
the results shown e.g. in Tables II-VI, X, XI, depending whether one takes the
exact $1/N$, IR approximated, or exact $N=2$ values of the relevant
perturbative coefficients $K_n$, given in Eq. (\ref{expNF}).

The large LDE order behavior, for $k\to\infty$, of expression
(\ref{ldep}) can also be analyzed, to give: 

\begin{equation}
\Delta \langle  \phi^2 \rangle^{(k)}(\eta,u) \raisebox{-0.4cm}{~\shortstack{ 
$\sim$ \\ $k\to\infty $}}
-\frac{N \eta}{4\pi} \frac{1}{\sqrt{\pi}\:k^{1/2}}
+\frac{u N}{3} \sum^k_{n=1} K_n
\left(-\frac{u N}{6 \eta}\right)^n\:  \frac{k^{n/2}}{\Gamma[1+n/2]}\;,
\label{ldepinf}
\end{equation}
where we used, to obtain Eq.~(\ref{ldepinf}), the well-known properties
 of the Gamma functions, such as
$\Gamma[2z] = 2^{2z-1/2} \Gamma[z] \, \Gamma[1/2+z]/\sqrt(2\pi)$, 
$\Gamma[z] \, \Gamma[1-z] \, \sin(\pi z)=\pi$ and the
Stirling asymptotic behavior $\Gamma[b + a z]\sim 
\sqrt{2\pi} e^{-a z}(a z)^{a z+b-1/2}$.
 
Expression (\ref{ldepinf}) clearly suggests to rescale for convenience the
arbitrary mass parameter according to $\eta \to\tilde \eta k^{1/2}$. Of course,
after  such a rescaling the relevant limit is again $\tilde \eta\to 0$. 
After such a rescaling, one
obtains 

\begin{equation}
\Delta \langle  \phi^2 \rangle^{(k\to\infty)}(\tilde \eta,u) \sim
-\frac{N \tilde \eta}{4\pi} \frac{1}{\Gamma[1/2]}
+\frac{u N}{3} \sum^\infty_{n=1} \frac{K_n}{\Gamma[1+n/2]}\:
\left(-\frac{u N}{6 \tilde \eta\;} \right)^n\:
\label{ldepinf2}
\end{equation}
which, as expected, agrees with the CIRT ``direct" LDE resummation
result Eq. (\ref{exp20cirt}).\\
More precisely, for the simpler geometric series cases Eq.~(\ref{simplde})
or Eq.~(\ref{serIR}), corresponding thus 
 to (up to an overall factor) $K_n =1$,
the sum in Eq.~(\ref{ldepinf2}) can be further
performed exactly, to give: $\exp(x^2)\:Erfc(x)-1$
with $x \equiv u/\eta$ (or $x \equiv u N/(48\pi\,\eta)$) for
Eq.~(\ref{simplde}) (respectively for Eq.~(\ref{serIR})). 
Finally the asymptotic expansion of $Erfc(x)$ for the relevant limit
$x\to\infty$ (equivalently $\eta\to 0$) \cite{abramowitz}:

\begin{equation} 
\label{Erfas}
\exp(x^2) Erfc(x) \sim
\frac{1}{\sqrt{\pi}\,x} \left[ 1+\sum^\infty_{q=1} \prod^q_{i=1}
\left (-\frac {1}{2x^2} \right )^q
(2i-1) \right]\;,
\end{equation}
was used at different stages in Sec. IV, for instance to examine
the behavior of the LDE series when the linear tadpole term is included in
the procedure.

\newpage

\begin{figure}[t]
\epsfysize=5.5cm
{\centerline{\epsfbox{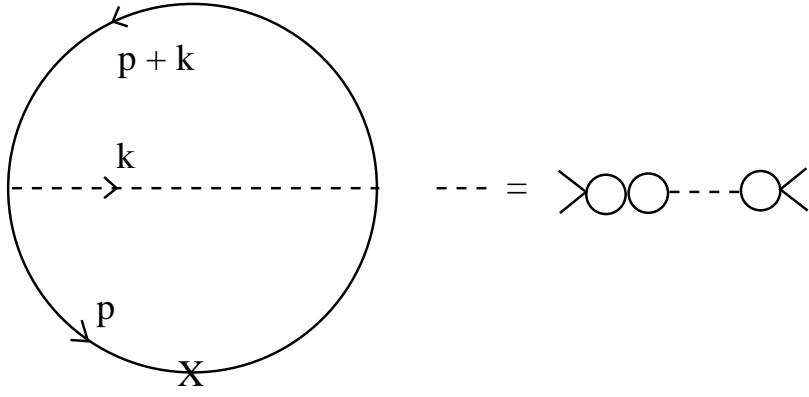}}}
\caption[]{\label{fig1} The Feynman graph for the relevant quantity
$\langle \phi^2 \rangle_u$ at $1/N$ order, with the resummed propagator
(dotted lines).}
\end{figure}


\begin{figure}[t]
\vspace{2cm}
\epsfysize=5.5cm
{\centerline{\epsfbox{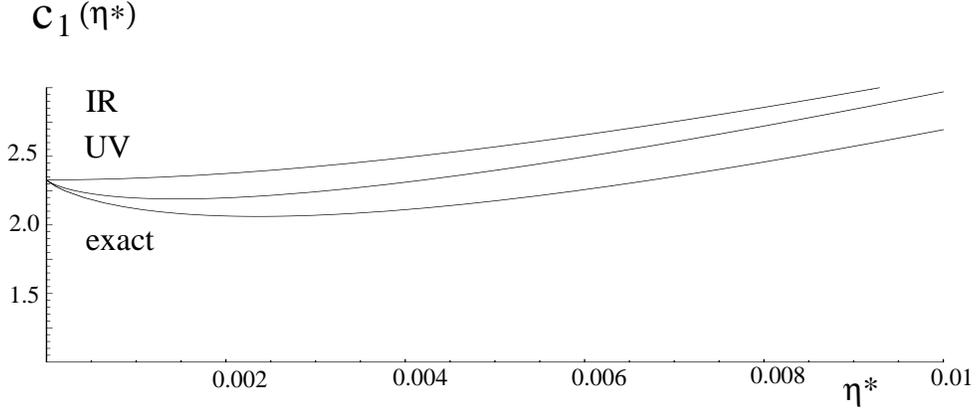}}}
\caption[]{\label{fig2} Comparison between the naive IR, UV propagator
approximation Eq. (\ref{asymN}), (\ref{asymUV}), and the exact numerical
integration of  Eq. (\ref{basic2N}), before subtracting the spurious
contribution $N u/(96\pi^2)$.}
\end{figure}


\begin{figure}[t]
\vspace{1cm}
\epsfysize=5.5cm
{\centerline{\epsfbox{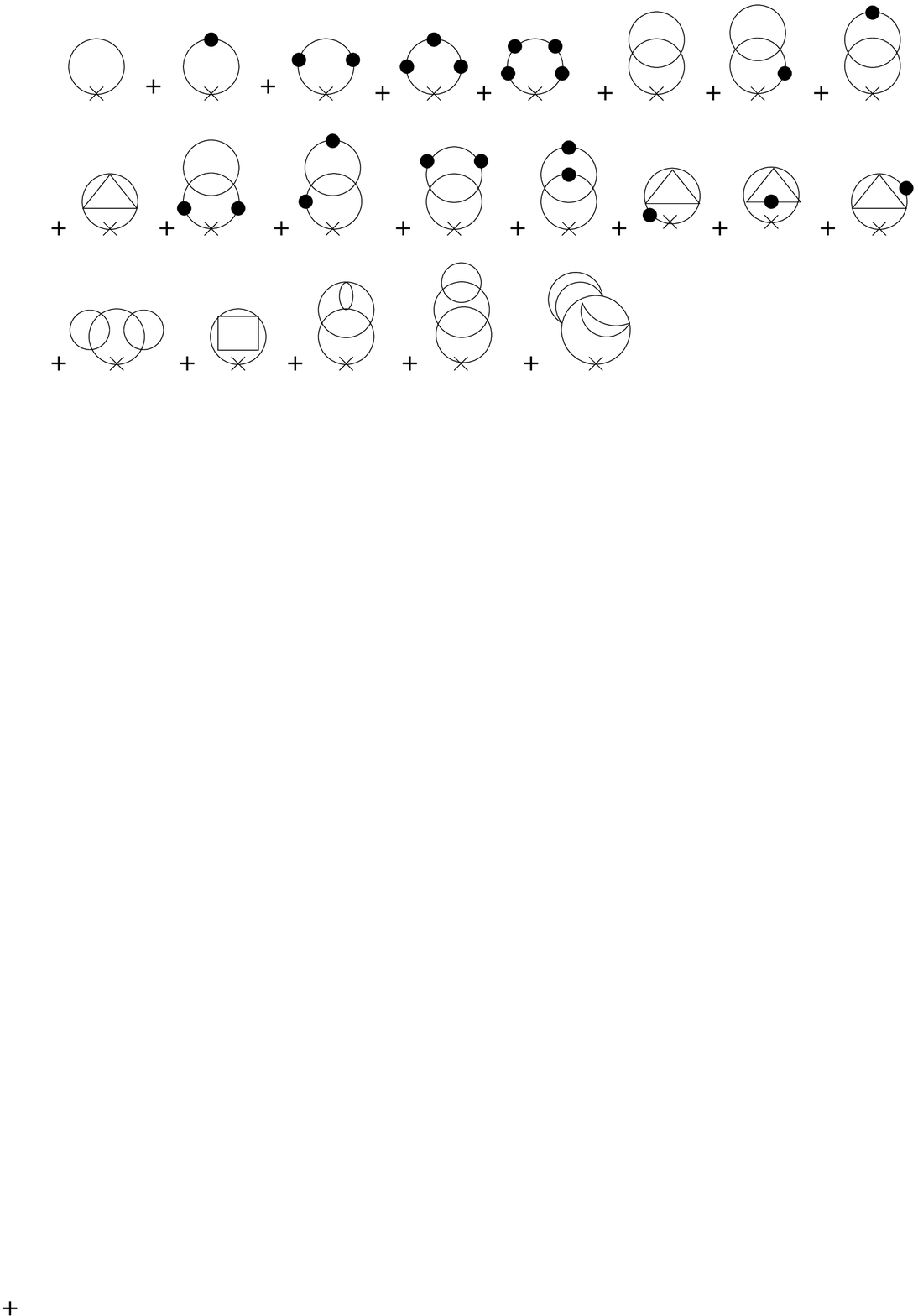}}}
\caption{All diagrams contributing to $\langle \phi^2 \rangle_u^{(4)}$
at the critical point. The black dot represents  the
$\delta \eta^2$ insertions.}
\end{figure}


\newpage

\begin{table}
\begin{center}
\caption{LDE of the simple alternated geometric series, Eq.(\ref
{simple}). PMS optimization results, at different
LDE orders $k$.}

\begin{tabular}{|c||c|c|c|c|c|c|}
\hline\hline
$ k $ & $S0$ & $S1$  &
$S2$ & $S3$ & $S4$ & $S5$  \\
\hline
\hline
1   & $\sqrt{2}$ & $-\sqrt{2}$ & -- & --&--&--\\
\hline
2   & 1.69743 & -1.12996 & -- & --&--&--\\
   &--&$ \pm 0.167336 I$& -- & -- &--&--\\
\hline
3  & 1.81088 &  -1.00296 & -1.11978 & -- &--&--\\
   &--&$ \pm 0.129109I$& -- & -- &--&--\\
\hline
4 &1.87082 &  -0.959082 &  -1.06361 & -- &--&-- \\
   &--&$\pm 0.076116 I$& $\pm 0.0465028I$ & -- &--&--\\
\hline
5 &1.90768& -0.950095 & -1.02417& -1.05508 &--&--  \\
   &--&$ \pm 0.03648 I$& $\pm 0.0499 I$ & -- &--&--\\
\hline
6 & 1.93258 &-0.954501& -1.00183  & -1.03612&--&--\\
    &--&$ \pm 0.01086 I$& $\pm 0.0407 I $ & $\pm 0.019 I$ &--&--\\
\hline
7 &1.95052& -0.9630 & -0.99069 & -1.01992 &-1.03154&-- \\
   &--&$ \pm 0.00447 I$&$\pm 0.02932 I$  &$\pm 0.02416 I$ & &--\\
\hline
8 &1.96405& -0.9719 &  -0.9861 &  -1.0085 & -1.02313 &--\\
   &--&$ \pm 0.01304 I$& $\pm 0.01929 I$ & $\pm 0.023 I $&$\pm 0.0096 I $&--\\
\hline
9 &1.97462&  -0.9797 &  -0.9851 & -1.0012 &  -1.01516 & -1.0205  \\
   &--&$ \pm 0.0174 I$&$\pm 0.0114 I$ &$\pm 0.0195 I$ &$\pm 0.01346 I$&--\\
\hline
10 &1.9831& -0.9861 & -0.9859 & -0.9968 & -1.0088 & -1.016\\
   &--&$ \pm 0.019 I$& $\pm 0.0056 I$ & $\pm 0.0154 I$&$\pm 0.0141I$&
$\pm 0.0055$ \\ \hline\hline
\end{tabular}
\end{center}
\end{table}

\begin{table}
\begin{center}
\caption{Standard LDE  at large-$N$, Eq. (\ref {exp20braaten}). PMS
results for $c_1$, at different orders $k$, obtained
with all families.}
\begin{tabular}{|c||c|c|c|c|c|c|c|c|c|c|c|}
\hline\hline
$ k $ & $F0$ &$ F1$  &  $F2$ & $F3$ & $F4$ & $F5$ & $F6$ & $F7$ & $F8$ & $F9$ & $F10$ \\
\hline \hline
2   & -2.163  &2.163&  -- & --  & --  & -- & -- & -- & --& --&-- \\
\hline
3  & -2.698 &1.879  & -- & -- & -- & -- & -- & -- & -- &-- & --\\
   & -- &$\pm 0.169 I$ & -- & -- & -- & -- & -- & -- &--& --&--\\
\hline
4 &-2.945  &$1.713 $ &$ 1.962 $ & -- & -- & --& -- & -- & -- & --&--\\
   &--& $\pm 0.061 I$ & -- & -- & -- & -- & -- & --&--& --&--\\
\hline
5 & -3.087&$ 1.642 $ & $1.913 $ &  -- & -- & -- & -- &-- &--& -- &--\\
   &--& $\pm 0.077 I$ &$ \pm 0.029 I$ & -- & -- & -- & -- & --& --& --&--\\
\hline
6 & -3.179&$ 1.620 $& $1.870 $& $1.935$ & -- & -- & -- & --& --&-- &--\\
   &--& $\pm 0.196 I$ &$ \pm 0.008 I$ & -- & -- & -- & -- & --& -- & --&--\\
\hline
7 & -3.244& $ 1.622 $ & $1.843 $ & $1.928 $ & -- & -- & -- & -- & -- & --&--\\
   &--& $\pm 0.293 I$ &$ \pm 0.067 I$ & $\pm 0.001 I$ & -- & -- & -- & -- & -- &-- &--\\
\hline
8 & -3.292&   $ 1.636 $     & $1.829 $      & $ 1.921 $     & 1.935 & -- & -- & -- & -- &-- &--\\
   &--& $\pm 0.369 I$ &$ \pm 0.128 I$ & $\pm 0.025 I$ & --    & -- & -- & -- & -- & --&--\\
\hline
9 & -3.329&   $ 1.654 $     & $1.825 $      & $1.913 $      &$ 1.936 $& $1.946$ & -- & -- & --& --&--\\
   &--& $\pm 0.429 I$ &$ \pm 0.184 I$ & $\pm 0.062 I$ & --      & --      & -- & -- & -- &-- &--\\
\hline
10 &-3.358 &  $1.675 $      &$1.828 $       & $1.908 $      &$1.938 $ &$ 1.950$     & -- & -- & -- &-- &--\\
   &--& $\pm 0.476 I$ &$ \pm 0.233 I$ & $\pm 0.102 I$ & --     & $\pm0.027 I$ & -- & -- & -- &-- &--\\
\hline
11 & -3.382&  $1.696 $      &$1.836 $       & $1.907 $      &$ 1.940$ &$ 1.950 $    & 1.966 & -- & -- &-- &--\\
   &--& $\pm 0.515 I$ &$ \pm 0.276 I$ & $\pm 0.139 I$ & --      &$0.056 I$    & --    & -- & -- &-- &--\\
\hline
12 &-3.402 &  $1.715 $      &$1.845 $       &$ 1.909 $      &$1.941 $ &$ 1.950 $       &$1.972$       & -- & -- &--& --\\
   &--& $\pm 0.546 I$ &$ \pm 0.313 I$ & $\pm 0.174 I$ & --      & $\pm 0.086 I $ &$\pm 0.025 I$ & -- & -- & --&--\\
\hline
13 & -3.419&  $1.733 $      &$1.856 $       &$ 1.913 $      &$1.942 $ &$ 1.950 $       & 1.974        & 1.983 & -- &-- &--\\
   &--& $\pm 0.571 I$ &$ \pm 0.344 I$ & $\pm 0.206 I$ & --      & $\pm 0.114 I $ & $0.050I$     & --& -- & --&--\\
\hline
14 & -3.433 & $1.750 $      & $1.868 $       & $ 1.919 $    & $1.942$ & $ 1.952 $      &$ 1.976 $      & 1.988         & --   &-- &--\\
   &--& $\pm 0.592 I$ & $ \pm 0.372 I$ & $\pm 0.234 I$& --      & $\pm 0.141 I $ & $\pm 0.075 I$ & $\pm 0.023I$  & -- & --  &--\\
\hline
15 & -3.445 & $1.765$       & $1.880$        & $1.926$      & $1.943$ & $1.956$        & $1.978 $      & $1.992$       & $1.997$ &-- &--\\
   &--& $\pm 0.610 I$ &$ \pm  0.395I$  & $\pm 0.260 I$& --      & $\pm  0.166I $ & $\pm 0.098I$  & $\pm 0.045I$  &     --  &-- &--\\
\hline
16 & -3.456 &  $1.780 $      & $1.891$        & $1.934$      & $1.944 $& $1.960$        & $1.981$       & $1.995$       & 2.002   &-- &--\\
   &--& $\pm 0.625 I$ &$ \pm 0.416 I$  & $\pm 0.282 I$ & --     & $\pm 0.189 I $ & $\pm 0.119 I$ & $\pm 0.066 I$ & $\pm 0.021 I$ &-- &--\\
\hline
17 & -3.466 & 1.792          & 1.903         & 1.942         & 1.944 & 1.966          & 1.984          & 1.998          & 2.006          & 2.010 &-- \\
   &--& $\pm  0.638I$ &$ \pm 0.433 I$ & $\pm 0.302 I$ & --    & $\pm 0.210 I $ & $ \pm 0.140 I$ & $\pm  0.085I $ & $\pm 0.040 I $ & -- & -- \\
\hline
18 &  -3.474 & 1.804         & 1.913         & 1.950         & 1.945 & 1.972          & 1.988         & 2.000         & 2.009          & 2.014 & --\\
   &--& $\pm 0.649 I$ &$ \pm 0.449 I$ & $\pm  0.321 I$& --    & $\pm 0.229 I $ & $\pm  0.159 I$& $\pm 0.104 I$ & $\pm 0.059 I$  & $\pm 0.019 I$ &--\\
\hline
19 &-3.482 & 1.815 & 1.924 & 1.959 & 1.945 & 1.978 & 1.992 & 2.004 & 2.012 & 2.018 & 2.020 \\
   &--& $\pm  0.658 I$ &$ \pm 0.463 I$ & $\pm 0.337 I  $ & -- & $\pm 0.246 I $ & $\pm 0.177 I$  & $\pm 0.121 I$ & $\pm 0.076  I$  & $\pm 0.036 I$ &--\\
\hline
20 &-3.488 & 1.825 & 1.933 & 1.967 & 1.945 & 1.984  & 1.997 & 2.007 & 2.016 & 2.021 & 2.025 \\
   &--& $\pm 0.666 I$ &$ \pm  0.475 I$ & $\pm 0.351 I$ & -- & $\pm 0.262 I $ & $\pm 0.193 I $  & $\pm 0.138 I $ & $\pm 0.092 I $  & $\pm 0.053 I $ & $\pm  0.017I $  \\
\hline\hline
\end{tabular}
\end{center}
\end{table}

\newpage

\begin{table}
\begin{center}
\caption{Standard LDE  at large-$N$, Eq.(\ref {exp20braaten}). 
Three families of
PMS and Fastest Apparent Convergence (FAC) results for $c_1$.}
\begin{tabular}{|c||c|c||c|c||c|c|}
\hline\hline
$ k $ & $ F1-PMS$  &  $F1-FAC$ &$F2-PMS$ &$F2-FAC$ & $F3-PMS$ &$F3-FAC$ \\
\hline \hline
2   &  2.163 & 2.498 & --  & --  & -- & -- \\
\hline
3  & $1.879\pm 0.169 I$ & $1.884 \pm 0.274 I$ & -- & -- & -- & -- \\
\hline
4 & $1.713 \pm 0.061 I$ & $1.660 \pm 0.097 I$ & $ 1.962 $ & 1.996 & -- & --\\
\hline
5 & $ 1.642 \pm 0.077 I$ & $1.593 \pm 0.078 I$ &$1.913 \pm 0.029 I$ & $1.917 \pm 0.0462 I$  & -- & --  \\
\hline
6 & $ 1.620 \pm 0.196 I$& $1.585 \pm 0.213 I$ & $1.870 \pm 0.008 I$& $1.858 \pm 0.003 I$ &$1.935$ & 1.941 \\
\hline
7 & $ 1.622 \pm 0.293 I$ & $1.600 \pm 0.314 I$ &$1.843 \pm 0.067 I$ & $1.826 \pm 0.063 I$ &$1.928 \pm 0.001 I$ & $1.928 \pm 0.005 I $\\
\hline
8 & $ 1.636 \pm 0.369 I$ & $1.622 \pm 0.390 I$  &$1.829 \pm 0.128 I$ & $1.813 \pm 0.130 I$  &$ 1.921  \pm 0.025 I$& $1.915 \pm 0.0213 I$\\
\hline
9 & $ 1.654 \pm 0.429 I$ & $1.647 \pm 0.448 I $ &$1.825 \pm 0.184 I$ & $1.811 \pm 0.189 I $ & $1.913 \pm 0.062 I$ & $ 1.905 \pm 0.061 I $\\
\hline
10 & $1.675 \pm 0.476 I$& $1.671 \pm 0.493 I$ &$1.828 \pm 0.233 I$& $1.817 \pm 0.241 I$  &$1.908 \pm 0.102 I$ & $1.899 \pm 0.102 I$\\
\hline
11 & $1.696 \pm 0.514 I$& $1.694 \pm 0.529 I$  &$1.836 \pm 0.276 I$& $1.827 \pm 0.284 I$  &$1.907 \pm 0.139 I$&  $1.898 \pm 0.141 I$ \\
\hline
12 & $1.715 \pm 0.545 I$& $1.715 \pm 0.558 I$ &$1.845 \pm 0.313 I$& $1.839 \pm 0.321 I$ &$ 1.908 \pm 0.174 I$&  $1.901 \pm 0.178 I$ \\
\hline
13 & $1.733 \pm 0.571 I$& $1.734 \pm 0.582 I$ &$1.856 \pm 0.344 I$& $1.852 \pm 0.353 I$  &$ 1.913 \pm 0.206 I$ & $1.906 \pm 0.210 I$ \\
\hline
14 & $1.750 \pm 0.592 I$& $1.752 \pm 0.601 I$ &$1.868 \pm 0.371 I$& $1.865 \pm 0.380 I $  &$ 1.919 \pm 0.234 I$ & $1.913 \pm 0.239 I $\\
\hline
15 & $1.765 \pm 0.610 I$& $1.767 \pm 0.618 I$  &$1.880 \pm 0.395 I$& $1.877 \pm 0.403 I$ &$ 1.926 \pm 0.260  I$&  $1.921 \pm 0.265 I $\\
\hline
16 & $1.780 \pm 0.625 I$& $1.782 \pm 0.631 I$ &$1.891 \pm 0.416 I$& $1.890 \pm 0.423 I$ &$ 1.934 \pm 0.282 I$& $ 1.930 \pm 0.287 I $\\
\hline
17 & $1.792 \pm 0.638 I$& $1.795 \pm 0.643 I $ &$1.903 \pm 0.433 I$& $1.902 \pm 0.440 I$ &$ 1.942 \pm 0.302 I$& $1.939 \pm 0.308 I$\\
\hline
18 & $1.804 \pm 0.649 I$& $1.807 \pm 0.654 I$ &$1.913 \pm 0.449 I$& $1.913 \pm 0.455 I$   &$ 1.950 \pm 0.321 I$& $1.947 \pm 0.326 I$ \\
\hline
19 & $1.815 \pm 0.658 I$& $1.817 \pm 0.663 I $ &$1.924 \pm 0.463 I$& $ 1.923 \pm 0.468 I$ &$ 1.959 \pm 0.337 I$& $ 1.957 \pm 0.342 I $\\
\hline
20 & $1.825 \pm 0.666 I$& $1.827 \pm 0.670 I$ &$1.933 \pm 0.475 I$& $1.934 \pm 0.480 I$ &$ 1.967 \pm 0.351 I $& $1.966 \pm 0.356 I$ \\
\hline\hline
\end{tabular}
\end{center}
\end{table}

\newpage

\begin{table}
\begin{center}
\caption{Infrared LDE  at large-$N$, Eq.(\ref {serIR}). 
PMS
results for $c_1$, at different orders $k$.}

\begin{tabular}{|c||c|c|c|c|c|c|c|c|c|c|c|}
\hline\hline
$ k $ & $F0$ & $ F1$  &
$F2$ & $F3$ & $F4$& $F5$&$F6$&$F7$&$F8$&$F9$&$F10$ \\
\hline
\hline
2   & -2.852 & 2.852 & -- & --&--&--&--&--&--&--&-- \\
\hline
3  & -3.577 & 2.444 & -- & -- &--&--&--&--&--&--&--\\
   &--&$ \pm 0.276 I$& -- & -- &--&--&--&--&--&--&--\\
\hline
4 &-3.910 & 2.244& 2.482 & -- &--&--&--&--&--&--&-- \\
   &--&$ \pm 0.200 I$& -- & -- &--&--&--&--&--&--&--\\
\hline
5 &-4.100& 2.184 & 2.397& -- &--&--&--&--&--&--&--  \\
   &--&$ \pm 0.097 I$& $\pm 0.079 I$ & -- &--&--&--&--&--&--&--\\
\hline
6 & -4.223 &2.184& 2.333  & 2.397&--&--&--&--&--&--&-- \\
    &--&$ \pm 0.020 I$& $\pm 0.081 I $ & -- &--&--&--&--&--&--&--\\
\hline
7 &-4.309& 2.205 & 2.298 & 2.369 &--&--&--&--&--&--&-- \\
   &--&$ \pm 0.028 I$&$\pm 0.060 I$  &$\pm 0.032 I$ &--&--&--&--&--&--&--\\
\hline
8 &-4.372& 2.232 & 2.283 & 2.342 & 2.366 &--&--&--&--&--&-- \\
   &--&$ \pm 0.055 I$& $\pm 0.037 I$ & $0.040 I $&--&--&--&--&--&--&--\\
\hline
9 &-4.214& 2.256 & 2.279 & 2.324 & 2.354 & -- &--&--&--&--&-- \\
   &--&$ \pm 0.068 I$&$\pm 0.016 I$ &$\pm 0.036 I$ &$\pm 0.016 I$&--&--&--&--&--&--\\
\hline
10 &-4.460& 2.277 & 2.282 & 2.313 & 2.341 & 2.352&--&--&--&--&--\\
   &--&$ \pm 0.074 I$& $\pm 0.010 I$ & $\pm 0.028 I$&$\pm 0.022 I$& --&--&--&--&--&--\\
\hline
11 &-4.498& 2.294&2.287  &2.307  &2.331  &2.346  & -- &--&--&--&--\\
   &--&$ \pm 0.074 I$& $\pm 0.010 I$ & $\pm 0.019 I$&$\pm 0.022 I$& $0.009 I$&--&--&--&--&--\\
\hline
12 & -4.518&2.307 & 2.293 & 2.305 & 2.324 & 2.339 & 2.345 &--&--&--&-- \\
   &--&$ \pm 0.072 I$& $\pm 0.017 I$ & $\pm 0.011 I$&$\pm 0.019 I$& $0.014 I$&--&--&--&--&--\\
\hline
13 & -4.540&  2.317 &2.299  &2.305  &2.319  &2.333  &2.341 & -- &-- &-- &-- \\
   &--&$ \pm 0.068 I$& $\pm 0.022 I$ & $\pm 0.004 I$&$\pm 0.015 I$& $0.015 I$&$\pm 0.006 I$&--&--&--&--\\
\hline
14 &-4.559 &2.324  &2.305 &2.306  &2.316 &2.328 &2.337 &2.340 &--&--&-- \\
   &--& $ \pm 0.064 I$& $\pm 0.025 I$ & $\pm 0.001 I$&$\pm 0.011 I$& $0.014 I$&$\pm 0.009 I$&--&--&--&--\\
\hline
15 &-4.575 &2.330 &2.310  &2.308  &2.315 &2.324 &2.333 &2.338 & --&-- &\\
   &--& $ \pm 0.061 I$& $\pm 0.027 I$ & $\pm 0.005 I$&$\pm 0.007 I$& $0.012 I$&$\pm 0.010 I$&$\pm 0.004 I$&--&--&--\\
\hline
16 &-4.589&2.334 &2.315 &2.310 &2.314 &2.322 &2.330 &2.335 &2.338 &--&--\\
   &--&$ \pm 0.057 I$& $\pm 0.027 I$ & $\pm 0.008 I$&$\pm 0.004 I$& $0.010 I$&$\pm 0.010 I$&$\pm 0.006 I$&--&--&--\\
\hline
17 &-4.602&2.338 &2.318 &2.313 &2.314 &2.320 &2.327 &2.333 &2.336 &--&--\\
   &--&$ \pm 0.053 I$& $\pm 0.027 I$ & $\pm 0.010 I$&$\pm 0.001 I$& $0.008 I$&$\pm 0.009 I$&$\pm 0.007 I$&$\pm 0.003 I$&--&--\\
\hline
18 &-4.613&2.340 &2.322 &2.315 &2.315 &2.319 &2.325 &2.330 &2.334 &2.336 &--\\
   &--&$ \pm 0.050 I$& $\pm 0.027 I$ & $\pm 0.012 I$&$\pm 0.001 I$& $0.005 I$&$\pm 0.008 I$&$\pm 0.007 I$&$\pm 0.004 I$& -- &--\\
\hline
19 &-4.623&2.342 &2.324 &2.317 &2.316 &2.319 &2.324 &2.329 &2.332 &2.335 &--\\
   &--&$ \pm 0.047 I$& $\pm 0.026 I$ & $\pm 0.013 I$&$\pm 0.003 I$& $0.003 I$&$\pm 0.007 I$&$\pm 0.007 I$&$\pm 0.005 I$& $\pm 0.002 I$ &--\\
\hline
20&-4.631&2.344&2.327&2.319&2.317&2.319&2.323&2.327&2.331&2.333&2.334\\
   &--&$ \pm 0.045 I$& $\pm 0.025 I$ & $\pm 0.014 I$&$\pm 0.005 I$& $0.002 I$&$\pm 0.005 I$&$\pm 0.006 I$&$\pm 0.007 I$&$\pm 0.003 I$&--\\
\hline\hline
\end{tabular}
\end{center}
\end{table}

\begin{table}
\begin{center}
\caption{Comparison of Infrared and exact LDE at large-$N$, 
for $\eta = R_c^{-1} = 1/(24\pi)$ at different orders $k$.}

\begin{tabular}{|c||c|c|c|c|c|c|c|c|c|c|c|}
\hline\hline
$ k $ & 1 & 2  & 3 & 4 & 5 & 6 & 7 & 8 & 9 & 10 &\\
\hline
IR & 1.16424 & 3.20165 & 1.89188 &
2.67411 &    2.17385 & 2.47173 & 2.27958 &
2.39236 &    2.31806 & 2.36035 & \\
\hline
exact & 1.16424& 2.2129& 1.78534& 1.98289& 1.892& 1.94579& 1.92615& 1.94536& 
 1.94367& 1.95312&  \\
\hline\hline
$ k $ & 11 & 12  & 13 & 14 & 15 & 16 & 17 & 18 & 19 & 20 & 100\\
\hline
IR & 
2.33129& 2.34685&    2.33524 & 2.34073&
2.33590& 2.33765 &    2.33549& 2.33588&
2.33481& 2.33474 &
2.32910 \\
\hline
exact &  1.95614& 1.9624& 1.96651& 1.9715& 1.97566& 1.97999& 
1.98393& 1.98782& 1.99149& 1.99504& 2.10987 \\
\hline\hline
\end{tabular}
\end{center}
\end{table}

\begin{table}
\begin{center}
\caption{Best PMS results for $c_1$, at different orders $k$, from the exact
LDE series at large-$N$ when omitting the tadpole term $-(N\eta)/(4\pi)$.}

\begin{tabular}{|c||c|c|c|c|c|c|c|c|c|c|c|}
\hline\hline
$ k $ & 3 & 4  & 5 & 6 & 10 & 15\\
\hline
$c_1$ & $1.061$ & $  1.222 \pm 0.37 I$ & $1.34 $ &
$1.435 \pm 0.720 I$ &  $1.710 \pm 0.99 I$ & $1.896 \pm 1.089 I$ \\
\hline\hline
$ k $ & 20 & 30 & 50 & 60 & 80 & 100 \\
\hline
$c_1$ & $1.999 \pm 1.120 I$ & $2.129 \pm 0.880 I$ & $2.243 \pm 0.900 I$
& $2.272 \pm 0.910 I$   & $2.311 \pm 0.780 I$ & $2.33 \pm 0.69 I$ \\
\hline\hline
\end{tabular}
\end{center}
\end{table}

\newpage

\begin{table}
\begin{center}
\caption{Standard LDE at large-$N$, Eq. (\ref {exp20cirt}). 
All CIRT-PMS
results for $c_1$, at different orders $k$.}
\begin{tabular}{|c||c|c|c|c|c|c|c|c|c|c|c|}
\hline\hline
$ k $ & $F0$& $ F1$  &
$F2$ & $F3$ & $F4$& $F5$&$F6$&$F7$&$F8$&$F9$&$F10$ \\
\hline
\hline
2  &-2.818 & 2.818 & -- & --&--&--&--&--&--&--&--\\
\hline
3  &-3.400& 2.000 & -- & -- &--&--&--&--&--&--&--\\
   &--    &$\pm 0.814 I$&--&--&--&--&--&--&--&--&-- \\
\hline
4 &-3.554& 1.219 & 2.428 & -- &--&--&--&--&--&--&-- \\
  &--    &$\pm 0.669 I$&--&--&--&--&--&--&--&-- &--\\
\hline
5 & -3.597&0.750 & 2.121& -- &--&--&--&--&--&--&--  \\
 &--      &$\pm 0.242 I$&$\pm 0.408 I$&--&--&--&--&--&--&--&--\\
\hline
6 & -3.610&0.506& 1.734  & 2.246&--&--&--&--&--&--&-- \\
 & --&$\pm 0.247 I$& $\pm 0.463 I$&--&--&--&--&--&--&--&--\\
\hline
7 & -3.613&0.416 & 1.422 & 2.096 &--&--&--&--&--&--&-- \\
 & --&$\pm 0.740 I$& $\pm 0.336 I$&$\pm 0.233 I$&--&--&--&--&--&--&--\\
\hline
8 & -3.614&0.448 & 1.205 & 1.882 & 2.147 &--&--&--&--&--&-- \\
 & --&$\pm 1.228 I$& $\pm 0.127 I$&$\pm 0.304 I$&--&--&--&--&--&--&--\\
\hline
9 & -3.614&0.587 & 1.071 & 1.681 & 2.062 & -- &--&--&--&--&-- \\
 & --&$\pm 1.691 I$& $\pm 0.115 I$&$\pm 0.269 I$&$\pm 0.146 I$&--&--&--&--&--&--\\
\hline
10 &-3.614&0.836 & 1.001 & 1.518 & 1.932 & 2.087&--&--&--&--&--\\
 & --&$\pm 2.138 I$& $\pm 0.365 I$&$\pm 0.171 I$&$\pm 0.206 I$&--&--&--&--&--&--\\
\hline
11 & -3.614&1.203& 0.998 & 1.397 & 1.797 & 2.034 & -- &--&--&--&--\\
 & --&$\pm 2.552 I$& $\pm 0.609 I$&$\pm 0.038 I$&$\pm 0.200 I$&$\pm 0.097 I$&--&--&--&--&--\\
\hline
12 & -3.614&1.701 & 1.040 & 1.317 & 1.678 & 1.949 & 2.047 &--&--&--&-- \\
 & --&$\pm 2.913 I$& $\pm 0.841 I$&$\pm 0.111 I$&$\pm 0.151 I$&$\pm 0.143 I$&--&--&--&--&--\\
\hline
13 & -3.614&2.343& 1.126 & 1.272 & 1.581 & 1.856 & 2.012& -- &-- &-- &-- \\
 & --&$\pm 3.190 I$& $\pm 1.054 I$&$\pm 0.266 I$&$\pm 0.073 I$&$\pm 0.148 I$&$\pm 0.067 I$&--&--&--&--\\
\hline
14 & -3.614&3.142 & 1.251& 1.58 &1.506&1.768&1.954&2.020&--&--&-- \\
 & --&$\pm 3.339 I$& $\pm 1.243 I$&$\pm 0.419 I$&$\pm 0.021 I$&$\pm 0.122 I$&$\pm 0.102 I$&--&--&--&--\\
\hline
15 &-3.614&4.106 & 1.412 & 1.272 &1.455&1.691&1.887&1.996&--&--&--\\
 & --&$\pm 3.301 I$& $\pm 1.401 I$&$\pm 0.566 I$&$\pm 0.125 I$&$\pm 0.073 I$&$\pm 0.110 I$&$\pm 0.048$&--&--&--\\
\hline
16 &-3.614&5.229&1.606&1.309&1.424&1.628&1.821&1.955&2.001&--&--\\
 & --&$\pm 2.996 I$& $\pm 1.525 I$&$\pm 0.702 I$&$\pm 0.232 I$&$\pm 0.010 I$&$\pm 0.095 I$&$\pm 0.074$&--&--&--\\
\hline
17 &-3.614&6.491&1.828&1.367&1.411&1.579&1.761&1.906&1.984&--&--\\
 & --&$\pm 2.327 I$& $\pm 1.604 I$&$\pm 0.825 I$&$\pm 0.339 I$&$\pm 0.062 I$&$\pm 0.063 I$&$\pm 0.082$&$\pm 0.035 I$&--&--\\
\hline
18 &-3.614&7.838 & 2.071&1.442&1.416&1.544&1.709&1.855&1.954&1.988&--\\
 & --&$\pm 1.169 I$& $\pm 1.632 I$&$\pm 0.932 I$&$\pm 0.442 I$&$\pm 0.140 I$&$\pm 0.019 I$&$\pm 0.073$&$\pm 0.055 I$&--&--\\
\hline
19 &-3.614&9.181& 2.327&1.532&1.434&1.522&1.665&1.807&1.917&1.975&--\\
 & --&$\pm 0.628 I$& $\pm 1.600 I$&$\pm 1.022 I$&$\pm 0.539 I$&$\pm 0.220 I$&$\pm 0.033 I$&$\pm 0.051$&$\pm 0.061 I$&$\pm 0.025 I$&--\\
\hline
20&-3.614& 10.360&2.584&1.634&1.466&1.513&1.632&1.764&1.877&1.952&1.978\\
 & --&$\pm 3.236 I$& $\pm 1.499 I$&$\pm 1.091 I$&$\pm 0.628 I$&$\pm 0.299 I$&$\pm 0.091 I$&$\pm 0.019$&$\pm 0.055 I$&$\pm 0.041 I$&--\\
\hline\hline
\end{tabular}
\end{center}
\end{table}

\newpage

\begin{table}
\begin{center}
\caption{Infrared LDE at large-$N$, Eq. (\ref {Erf}). All CIRT-PMS large-$N$ 
results for $c_1$, at different orders $k$.}

\begin{tabular}{|c||c|c|c|c|c|c|c|c|c|}
\hline\hline
$ k $ & $ F1$  &
$F2$ & $F3$ & $F4$& $F5$&$F6$&$F7$&$F8$&$F9$\\
\hline
\hline
2   & 2.328 & -- & --&--&--&--&--&--&--\\
\hline
3  & 2.328 & 2.262& 2.395 &--&--&--&--&--&--\\
\hline
4 & 2.328& 2.320 & 2.337 &--&--&--&--&--&--\\
   & -- & $\pm 0.067 I$ & $\pm 0.067 I$ &--&--&--&--&--&--\\
\hline
5 & 2.328 & 2.369& 2.287 &2.271&2.386&--&--&--&--\\
   &--& $\pm 0.054 I$ & $\pm 0.054 I$ &--&--&--&--&--&--\\
\hline
6 & 2.328& 2.389  & 2.268&2.294&2.363&--&--&--&--\\
    &--& $\pm 0.027 I $ & $\pm 0.027 I$&$\pm 0.041 I$&$\pm 0.041 I$&--&--&--&--\\
\hline
7 & 2.328 & 2.391 & 2.266 &2.323&2.334&2.281&2.376&--&--\\
   &--&$\pm 0.005 I$  &$\pm 0.005 I$ &$\pm 0.050 I$&$ \pm 0.050 I$&--&--&--&--\\
\hline
8 & 2.328 & 2.386 & 2.270 & 2.344 &2.313&2.293&2.364&--&-- \\
   &--& $\pm 0.011 I$ & $0.011 I $&$\pm 0.045 I$&$\pm 0.045 I$&$\pm 0.025 I$& $\pm 0.025 I$&--&--\\
\hline
9 & 2.328 & 2.379 & 2.278 & 2.357 & 2.300 &2.310&2.347&2.289&2.368\\
   &--&$\pm 0.021 I$ &$\pm 0.021 I$ &$\pm 0.036 I$&$\pm 0.036 I$&$\pm 0.036 I$&$\pm 0.036 I$&--&--\\
\hline
10 & 2.328 & 2.371 & 2.286 & 2.363 & 2.294&2.325&2.332&2.296&2.360\\
   &--& $\pm 0.027 I$ & $\pm 0.027 I$&$\pm 0.025 I$&$\pm 0.025 I$&$\pm 0.038 I$&$\pm 0.038 I$&$\pm 0.017 I$& $\pm 0.017 I$\\
\hline\hline
\end{tabular}
\end{center}
\end{table}

\begin{table}
\begin{center}
\caption{PMS optimization of Pad\'e approximants (based both on the original
and CIRT improved  series for two of the relevant families)
for $c_1$ in  the large $N$ case with the exact series coefficients.}

\begin{tabular}{|c||c|c||c|c||c|c||c|c|}
\hline\hline
$ k $ & $P[1,2]$ PMS & & $P[1,2]$ CIRT & & $P[0,3]$ PMS & & $P[0,3]$ CIRT & \\ \hline
\hline
3   & 2.623 & 2.034 &  2.01 $\pm$ 0.14 I& -- & 2.328 $\pm$ 0.18 I & --&2.328
$\pm$ 0.05I & -- \\ \hline
4  &2.61 & 2.05 & 1.877 $\pm$ 0.34 I & 2.02 & 2.328 $\pm$ 0.17 I& --& 2.328
$\pm$ 0.05I& -- \\ \hline
5 & 2.615 $\pm$ 0.10 I &  2.04  $\pm$ 0.1 I& 1.77 $\pm$ 0.61 I & 1.98 $\pm$ 0.06 I   
& 2.29 $\pm$ 0.2 I& 2.37 $\pm$ 0.19 I&
2.34 $\pm$ 0.05I &2.33 $\pm$ 0.05 I \\ \hline\hline
\end{tabular}
\end{center}
\end{table}

\newpage

\begin{table}
\begin{center}
\caption{All ordinary PMS
and CIRT-improved  PMS optimization 
for $c_1$ in  the finite $N=2$ case.}

\begin{tabular}{|c||c|c||c|c||c|c|}
\hline\hline
$ k $ & $F0$ PMS & $F0$ CIRT & $F1$ PMS & $F1$ CIRT & $F2$ PMS & $F2$ CIRT\\
\hline
\hline
2   & -3.05916   & -3.98590 & 3.05916  & 3.98590 &--&-- \\
\hline
3  &-4.47035  & - 5.97078 & 2.44730 & 3.10543 &-- &--  \\
    &--&--       &$\pm 1.65256 I$&$\pm 3.09300 I$&--&--\\
\hline
4 & -5.30592 &  -7.03900 & 1.53443 & 1.19134 & 3.14286  & 5.22847\\
    &--&--       &$\pm 2.29581 I$&$\pm 4.33683 I$&--&--\\
\hline\hline
\end{tabular}
\end{center}
\end{table}

\begin{table}
\begin{center}
\caption{Same as previous table
for $c_1$ in  the finite $N=2$ case, but with 
IR large $N$ perturbative coefficient $K_i$ for $i >3$ }

\begin{tabular}{|c||c|c||c|c||c|c|}
\hline\hline
$ k $ & $F0$ PMS & $F0$ CIRT & $F1$ PMS & $F1$ CIRT & $F2$ PMS & $F2$ CIRT\\
\hline
\hline
2   & -3.059   & -3.986 & 3.059  & 3.986 &--&-- \\
\hline
3  &-4.470  & - 5.971 & 2.447 & 3.105 &-- &--  \\
    &--&--       &$\pm 1.653 I$&$\pm 3.093 I$&--&--\\
\hline
4 & -5.306 &  -7.039 & 1.534 & 1.194 & 3.143  & 5.106\\
    &--&--       &$\pm 2.296 I$&$\pm 4.337 I$&--&--\\
\hline
5 & -5.717 & -7.05 & 1.352 & 1.176 & 3.71  & 5.09\\
    &--&--       &$\pm 2.83 I$&$\pm 4.33 I$&--&--\\
\hline
6 & -5.97 &  -7.05 & 1.29 & 1.179 & 4.00  & 5.09\\
    &--&--       &$\pm 3.13 I$&$\pm 4.32 I$&--&--\\
\hline
10 & -6.43 & -7.05 & 1.219 & 1.179 & 4.49  & 5.09\\
    &--&--       &$\pm 3.66 I$&$\pm 4.32 I$&--&--\\
\hline\hline
\end{tabular}
\end{center}
\end{table}

\begin{table}
\begin{center}
\caption{PMS optimization of Pad\'e approximants (based both on the original
and CIRT improved  series for two of the relevant families)
for finite $N=2$ case.}

\begin{tabular}{|c||c|c||c|c||c|c||c|c|}
\hline\hline
$ k $ & $P[1,2]$ PMS & & $P[1,2]$ CIRT & & $P[0,3]$ PMS & & $P[0,3]$ CIRT & \\ \hline
\hline
3   & 2.62 &2.04 &  2.01 $\pm$ 0.20 I& -- & 2.328 $\pm$ 3.24 I & --&1.37
 & 3.28 \\ \hline
4  &2.60 & 2.06 & 1.98 $\pm$ 0.41 I & 1.98 & 2.328 $\pm$ 3.07 I& --& 0.97
$\pm$ 0.34 I& 3.15 \\ \hline
5 & 2.63 $\pm$ 0.07 I &  2.02  $\pm$ 0.07 I& 1.97 $\pm$ 0.62 I & 1.94 $\pm$
0.12 I    & 1.51 $\pm$ 3.45 I& 3.15 $\pm$ 3.45 I&
1.11 $\pm$ 0.44 I &3.17 \\ \hline
10 & 2.60 $\pm$ 0.04 I &  2.05  $\pm$ 0.04 I& 1.82 $\pm$ 0.31 I & 1.88 $\pm$
0.11 I    & 1.20 $\pm$ 1.65 I& 3.45 $\pm$ 1.65 I&
1.15 $\pm$ 0.11 I &3.27\\ \hline\hline
\end{tabular}
\end{center}
\end{table}

\end{document}